%% file: main.tex
\documentclass{article}

     \usepackage[nonatbib,final]{neurips_2020}

\usepackage[utf8]{inputenc} 
\usepackage[T1]{fontenc}    
\usepackage{hyperref}       
\hypersetup{colorlinks=true,linkcolor=blue,citecolor=red}
\usepackage{url}            
\usepackage{booktabs}       
\usepackage{amsfonts}       
\usepackage{nicefrac}       
\usepackage{microtype}      
\usepackage{amsmath}
\usepackage{amsthm}
\usepackage[ruled]{algorithm2e}
\usepackage{algorithmic}
\usepackage{xcolor}
\usepackage{graphicx}
\usepackage{subfig}

\definecolor{green}{RGB}{0, 100, 0}

\input{commands.tex}

\title{A biologically plausible neural network for \\ Slow Feature Analysis}

\author{David Lipshutz\thanks{Equal contribution}\hspace{3pt}$^{\,1}$\hspace{20pt}Charlie Windolf\protect\footnotemark[1]\hspace{3pt}$^{\,1,2}$\hspace{20pt}Siavash Golkar$^{\,1}$ \hspace{20pt} Dmitri B.\ Chklovskii$^{\,1,3}$  
  \vspace{14pt}
   \\
   $^{1\,}$Center for Computational Neuroscience, Flatiron Institute
   \\
   $^{2\,}$Department of Statistics, Columbia University
   \\
   $^{3\,}$Neuroscience Institute, NYU Medical Center
   \vspace{10pt}\\
   \texttt{\{dlipshutz,sgolkar,dchklovskii\}@flatironinstitute.org} \\
   \texttt{c.windolf@columbia.edu}
}

\begin{document}

\maketitle

\begin{abstract}
Learning latent features from time series data is an important problem in both machine learning and brain function. 
One approach, called Slow Feature Analysis (SFA), leverages the slowness of many salient features relative to the rapidly varying input signals.
Furthermore, when trained on naturalistic stimuli, SFA reproduces interesting properties of cells in the primary visual cortex and hippocampus, suggesting that the brain uses temporal slowness as a computational principle for learning latent features.
However, despite the potential relevance of SFA for modeling brain function, there is currently no SFA algorithm with a biologically plausible neural network implementation, by which we mean an algorithm operates in the online setting and can be mapped onto a neural network with local synaptic updates.
In this work, starting from an SFA objective, we derive an SFA algorithm, called Bio-SFA, with a biologically plausible neural network implementation. 
We validate Bio-SFA on naturalistic stimuli.
\end{abstract}

\section{Introduction}

Unsupervised learning of meaningful latent features from noisy, high-dimensional data is a fundamental problem for both machine learning and brain function.
Often, the relevant features in an environment (e.g., objects) vary on relatively slow timescales when compared to noisy sensory data (e.g., the light intensity measured by a single receptor in the retina).
Therefore, temporal slowness has been proposed as a computational principle for extracting relevant latent features \cite{Fldik1991,mitchison1991removing,wiskott2002slow}.

A popular approach for extracting slow features, introduced by Wiskott and Sejnowski \cite{wiskott2002slow}, is Slow Feature Analysis (SFA).
SFA is an unsupervised learning algorithm that extracts the slowest projection, in terms of discrete time derivative, from a nonlinear expansion of the input signal.
When trained on natural image sequences, SFA extracts features that resemble response properties of complex cells in early visual processing~\cite{berkes2005slow}.
Impressively, hierarchical networks of SFA trained on simulated rat visual streams learn representations of position and orientation similar to representations encoded in the hippocampus~\cite{franzius2007hsfa}.

The relevance of SFA is strengthened by its close relationship to information theoretic objectives and its equivalence to other successful algorithms under certain assumptions.
When the time series is reversible and Gaussian, (Linear) SFA is equivalent to maximizing mutual information between the current output of the system and the next input \cite{creutzig2008predictive,clark2019unsupervised}. 
Moreover, features extracted by several algorithms favoring predictability from real-world datasets are similar to those extracted by SFA \cite{weghenkel2018slowness}.
Finally, (Linear) SFA is equivalent to a time-lagged independent components analysis \cite{BlaschkeBerkesEtAl-2006,Hyvrinen2000}, which is a popular statistical technique used to analyze molecular dynamics \cite{PrezHernndez2013,No2015,Schwantes2015,MSultan2017}.

Due to its success in modeling aspects of neural processing, deriving an algorithm for SFA with a biologically plausible neural network implementation is an important task.
For the purposes of this work, we define biologically plausible to mean that the neural network operates in the online setting (i.e., after receiving an input, it computes its output before receiving its next input, never storing a significant fraction of past inputs), and its synaptic learning rules are local (i.e., a synaptic weight update depends only on variables represented in the pre- and postsynaptic neurons).
In addition to satisfying basic properties of neural circuits, these online and locality requirements can lead to networks that are well-suited for analyzing large datasets because they operate in the online setting with low computational overhead.

While there are a few online algorithms for SFA, none have biologically plausible neural network implementations that extract multiple slow features.
Moreover, there are no neural network implementations for the related information theoretic algorithms discussed above \cite{weghenkel2018slowness,clark2019unsupervised}.
Kompella et al.\ propose Incremental SFA \cite{kompella2012incremental} (see \cite{liwicki2012iksfa,Yousefi2012} for extensions). 
However, this approach relies on non-local learning rules, so it does not meet the above criteria for biological plausibility.
Malik et al.\ \cite{Malik2014} use an online generalized eigenvalue problem solver \cite{QingfuZhang2000} to derive an online algorithm for SFA.
While their algorithm for finding one-dimensional projections can be implemented in a biologically plausible network, their extension to multi-dimensional projections is not fully online.

\begin{figure}
\centering
\makebox[0pt][c]{\parbox{\textwidth}{%
    \begin{minipage}[c]{0.5\textwidth}\centering
    \resizebox{1\textwidth}{!}{
        \includegraphics{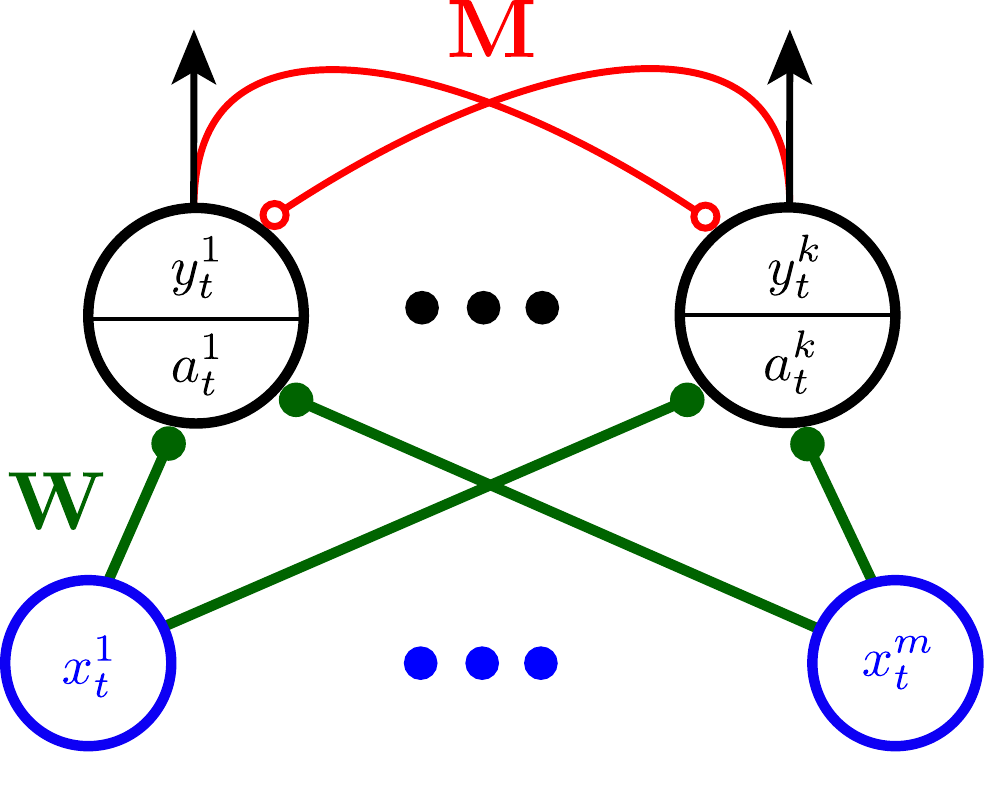}
        }
    \end{minipage}
    \hfill
    \begin{minipage}[c]{0.46\textwidth}\centering
    \resizebox{1\textwidth}{!}{
        \renewcommand{\arraystretch}{1.25}
        \small
        \begin{tabular}{  c  c }
            \toprule
            Variable & Biological interpretation \\
            \midrule
            ${\color{blue}\x_t}$ & expanded signal \\ 
            ${\color{green}\W}$ & feedforward synaptic weights \\ 
            $\a_t:={\color{green}\W}{\color{blue}\x_t}$ & dendritic current \\ 
            ${\color{red}\M}$ & lateral synaptic weights \\ 
            $\y_t$ & output signal
            \vspace{5pt}
            \\ 
            \toprule
            \multicolumn{2}{c}{Neural dynamics \& plasticity rules}\\
            \midrule
            \multicolumn{2}{l}{$d\y_t(\gamma)/d\gamma=\a_t-{\color{red}\M}\y_t(\gamma)$}\vspace{3pt}\\ 
            \multicolumn{2}{l}{$\Delta{\color{green}\W}=2\eta((\y_t+\y_{t-1})({\color{blue}\x_t}+{\color{blue}\x_{t-1}})^\top-\a_t{\color{blue}\x_t}^\top)$}\vspace{3pt}\\
            \multicolumn{2}{l}{$\Delta{\color{red}\M}=\frac{\eta}{\tau}((\y_t+\y_{t-1})(\y_t+\y_{t-1})^\top-{\color{red}\M})$}
            \vspace{3pt}\\
        \bottomrule
        \end{tabular}
        }
    \end{minipage}%
}}
\caption{A biologically plausible neural network implementation of Bio-SFA. The figure on the left depicts the architecture of the neural network. Blue circles are the input neurons and black circles are the output neurons with separate dendritic and somatic compartments. Lines with circles connecting the neurons denote synapses. Filled (resp.\ empty) circles denote non-Hebbian (resp.\ anti-Hebbian) synapses.}
\label{fig:NN}
\end{figure}

In this work, we propose Bio-SFA: an online algorithm for SFA with a biologically plausible neural network implementation, Fig.~\ref{fig:NN}.
We adopt a normative approach to derive our algorithm.
First, we express the solution of the SFA problem in terms of an objective from classical multidimensional scaling.
We then manipulate the objective to arrive at a min-max optimization problem that can be solved in the online setting by taking stochastic gradient descent-ascent steps.
These steps can be expressed in terms of neural activities and updates to synaptic weight matrices, which leads to a natural interpretation of our online algorithm as a biologically plausible neural network.
To validate our approach, we test our algorithm on datasets of naturalistic stimuli and reproduce results originally performed in the offline setting.

The synaptic updates of the feedforward weights $\W$ in our network are similar,  although not identical, to the updates proposed heuristically by F\"oldi\'ak~\cite{Fldik1991} to extract slow temporal features. 
However, there is no theoretical analysis of the algorithm in \cite{Fldik1991}. 
In contrast, in our normative approach, Bio-SFA is derived directly from an SFA objective, so we can analytically predict its output, as well as the synaptic weights, without resorting to numerical simulation. 
In addition, the comparison of our learning rules with F\"oldi\'ak's illuminates the relationship of \cite{Fldik1991} to SFA.

\section{Slow Feature Analysis}
\label{sec:sfa}

Here and below, vectors are boldface lowercase letters  (e.g., $\v$), and matrices are boldface uppercase letters (e.g., $\M$).
We use superscripts to denote the components of a vector (e.g., $v^i$).

\subsection{Problem statement}

Wiskott and Sejnowski \cite{wiskott2002slow} proposed the following 2 step method for extracting slow features from a noisy data set: (1) generate a nonlinear expansion of the input signal, and (2) find the slowest, in terms of discrete time derivative, low-dimensional projection of the expanded signal.
In this section, we review these 2 steps.

Let $\{\s_0,\s_1,\dots,\s_T\}$ be a $d$-dimensional input signal.\footnote{The zeroth time step is included to ensure the discrete-time derivative is defined at $t=1$.}
The first step of SFA is to generate an $m$-dimensional expansion $\{\x_t\}$, referred to as the expanded signal, of $\{\s_t\}$.
Let ${\bf h}=(h^1,\dots,h^m):\R^d\to\R^m$ be an expansion function and define
\begin{align*}
    \x_t:={\bf h}(\s_t)-\frac1T\sum_{t'=1}^T{\bf h}(\s_{t'}),\qquad t=0,1,\dots,T,
\end{align*}
so that $\{\x_t\}$ is centered.

Let $k<m$. 
The second step of SFA is to find the $k$-dimensional linear projection $\{\y_t\}$ of the expanded signal $\{\x_t\}$ that minimizes the mean discrete-time derivative of the output signal $\{\y_t\}$, subject to a whitening constraint.
To be precise, the objective can be formulated as follows:
\begin{align}
\label{eq:sfa1}
    \argmin{\{\y_t\}}\frac1T\sum_{t=1}^T\|\dot\y_t\|^2\quad\text{subject to}\quad\frac1T\sum_{t=1}^T\y_t\y_t^\top=\I_k,
\end{align}
where $\dot\y_t$ is the discrete time derive of $\y_t$, and $\y_t$ is a linear projection of $\x_t$; that is,
\begin{align}
    \label{eq:sfa2}
    \dot\y_t&:=\y_t-\y_{t-1},&& t=1,\dots,T,\\ 
    \label{eq:sfa3}
    \y_t&:=\V^\top\x_t,&& t=0,1,\dots,T,\qquad\text{ for some }\V\in\R^{m\times k}.
\end{align}
Note, since $\{\x_t\}$ is centered, the projection $\{\y_t\}$ is also centered.

\subsection{Quadratic SFA}
\label{sec:qSFA}

The focus of this work is to derive a biologically plausible neural network that learns to output the optimal output signal $\{\y_t\}$ when streamed the expanded signal $\{\x_t\}$.
While our algorithm does not depend on the specific choice of the expansion function ${\bf h}$, for concreteness, we provide an example here.

In their original paper, Wiskott and Sejnowski \cite{wiskott2002slow} proposed setting the components of the function ${\bf h}:\R^d\to\R^m$ to be the monomials of degree one and two.
This choice, which we refer to as ``Quadratic SFA'', has been widely used in applications \cite{wiskott2002slow,berkes2005slow,franzius2007hsfa,zhang2012slow}.
In particular, let $m:=d+d(d+1)/2$ and $h^1,\dots,h^m:\R^d\to\R$ denote the $m$ possible linear and quadratic functions of the form
\begin{align*}
    h(\s):=s^i\qquad\text{or}\qquad h(\s):=s^is^j,
\end{align*}
for $1\le i\le j\le d$.
(When only the linear features are used, i.e., $x^i=s^i+\text{const}$, this is referred to ``Linear SFA''.)
Thus, each component of the output signal is a quadratic polynomial in the components of the signal of the form:
\begin{align}
\label{eq:quadpoly}
    y^i=V_{1i}h^1(\s)+\cdots+V_{mi}h^m(\s)+\text{const}.
\end{align}
Biologically, there are a number of mechanism that have been proposed for computing products of the form $s^is^j$; see, e.g., \cite{koch1992multiplying} and the references therein.
One such mechanism uses ``Sigma-Pi'' units \cite{rumelhart1986general}, which multiplies two inputs via gating and have been invoked in cortical modeling \cite{mel1990sigma}.

In Sec.~\ref{sec:numerics}, we perform our numerical experiments using the quadratic expansion.

\section{A novel SFA objective from classical multidimensional scaling}

To derive an SFA network, we identify an objective function whose optimization leads to an online algorithm that can be implemented in a biologically plausible network.
To identify the objective function, we first rewrite the SFA output as a principal subspace projection and then take advantage of the fact that principal subspace projections can be expressed as solutions of objectives from classical multidimensional scaling \cite{cox2000multidimensional}.
This approach is similar to the derivation of a biologically plausible neural network for canonical correlation analysis \cite{lipshutz2020biologically}. 

To begin, we define the discrete derivative process $\{\dot\x_t\}$ and the delayed sum process $\{\bar\x_t\}$ by $\dot\x_t:=\x_t-\x_{t-1}$ and $\bar\x_t:=\x_t+\x_{t-1}$, for $t=1,\dots,T$.
In addition, we define the sample covariance matrices
\begin{align}
\label{eq:cov}
    \C_{xx}:=\frac1T\sum_{t=1}^T\x_t\x_t^\top,&&\C_{\dot x\dot x}:=\frac1T\sum_{t=1}^T\dot\x_t\dot\x_t^\top,&&\C_{\bar x\bar x}:=\frac1T\sum_{t=1}^T\bar\x_t\bar\x_t^\top.
\end{align}
Substituting the definitions in Eqs.~\eqref{eq:sfa2}, \eqref{eq:sfa3} and \eqref{eq:cov} into the objective in Eq.~\eqref{eq:sfa1}, we can equivalently write the SFA problem as the following constrained minimization problem of the projection matrix~$\V$:
\begin{align}
\label{eq:min}
    \argmin{\V\in\R^{m\times k}}\tr\V^\top\C_{\dot x\dot x}\V\quad\text{subject to}\quad\V^\top\C_{xx}\V=\I_k.
\end{align}
Due to the whitening constraint in Eq.~\eqref{eq:min}, we can equivalently write it as the maximization of the one-step autocorrelation of the projection $\{\y_t\}$ (see Appendix \ref{apdx:derivation} for details):
\begin{align}
\label{eq:genEV}
    \argmax{\V\in\R^{m\times k}}\tr\V^\top\C_{\bar x\bar x}\V\quad\text{subject to}\quad\V^\top\C_{xx}\V=\I_k.
\end{align}

Next, setting $\hat\x_t:=\C_{xx}^{-1/2}\bar\x_t$ for $t=1,\dots,T$, and
\begin{align*}
    \hat\V:=\C_{xx}^{1/2}\V,&&\C_{\hat x\hat x}:=\frac1T\sum_{t=1}^T\hat\x_t\hat\x_t^\top=\C_{xx}^{-1/2}\C_{\bar x\bar x}\C_{xx}^{-1/2},
\end{align*}
we see that $\V$ is a solution of Eq.~\eqref{eq:genEV} if and only if $\hat\V$ is the solution of:
\begin{align}
    \label{eq:pca}
    \argmax{\hat\V\in\R^{m\times k}}\tr\hat\V^\top\C_{\hat x\hat x}\hat\V\quad\text{subject to}\quad\hat\V^\top\hat\V=\I_k.
\end{align}
Notably, Eq.~\eqref{eq:pca} is the variance maximization objective for the PCA eigenproblem, which is optimized when the column vectors of $\hat \V$ span the $k$-dimensional principal subspace of $\C_{\hat x\hat x}$.

Finally, we take advantage of the fact that principal subspace projections can be expressed as solutions of objectives from classical multidimensional scaling \cite{cox2000multidimensional,pehlevan2015normative}. 
To this end, define the data matrices
\begin{align*}
    \bar\X:=[\bar\x_t,\dots,\bar\x_T],&&\hat\X:=[\hat\x_1,\dots,\hat\x_T],&&\bar\Y:=[\bar\y_1,\dots,\bar\y_T].
\end{align*}
Then, since $\bar\y_t=\V^\top\bar\x_t=\hat\V^\top\hat\x_t$, we see that $\bar\Y$ is the projection of $\hat\X_t$ onto its $k$-dimensional principal subspace.
As shown in \cite{cox2000multidimensional}, this principal projection can be expressed as a solution of the following objective from classical multidimensional scaling:
\begin{align}
\label{eq:similarity}
  \argmin{\bar\Y\in\R^{k\times T}}\frac{1}{2T^2}\big\|\bar\Y^\top\bar\Y-\hat\X^\top\hat\X\big\|_\text{Frob}^2=\argmin{\bar\Y\in\R^{k\times T}}\frac{1}{2T^2}\big\|\bar\Y^\top\bar\Y-\bar\X^\top\C_{xx}^{-1}\bar\X\big\|_\text{Frob}^2.
\end{align}
This objective minimizes the difference between the similarity of consecutive sums of output pairs, $\bar\y_t^\top\bar\y_{t'}$, and the similarity of consecutive sums of whitened input pairs, $\hat\x_t^\top\hat\x_{t'}$, where similarity is measured in terms of inner products.
Here we have assumed that $\C_{xx}$ is full rank.
If $\C_{xx}$ is not full rank (but is at least rank $k$), we can replace $\C_{xx}^{-1}$ in Eq.~\eqref{eq:similarity} with the Moore-Penrose inverse $\C_{xx}^+$ (see Appendix \ref{apdx:derivation}).

\section{Derivation of an online algorithm}
\label{sec:alg}

While the objective \eqref{eq:similarity} can be minimized by taking gradient descent steps in $\bar\Y$, this does not lead to an online algorithm because the gradient steps require combining inputs from different time steps.
Instead, we rewrite the objective as a min-max problem that can be solved by taking gradient descent-ascent steps that correspond to neural activities and synaptic update rules.

\subsection{A min-max formulation}

Expanding the square in Eq.~\eqref{eq:similarity} and dropping terms that do not depend on $\bar\Y$, we obtain the minimization problem
\begin{align}
\label{eq:minimization}
    &\min_{\bar\Y\in\R^{k\times T}}\frac{1}{2T^2}\tr\left(\bar\Y^\top\bar\Y\bar\Y^\top\bar\Y-2\bar\Y^\top\bar\Y\bar\X^\top\C_{xx}^{-1}\bar\X\right).
\end{align}
By introducing dynamical matrix variables $\W$ and $\M$, which will correspond to synaptic weights, we can rewrite the minimization problem \eqref{eq:minimization} as a min-max problem:
\begin{align*}
    \min_{\bar\Y\in\R^{k\times T}}\min_{\W\in\R^{k\times n}}\max_{\M\in\mathcal{S}_{++}^k}L(\W,\M,\bar\Y),
\end{align*}
where $\mathcal{S}_{++}^k$ denotes the set of $k\times k$ positive definite matrices and
\begin{align}
\label{eq:L}
    L(\W,\M,\bar\Y):=\frac{1}{T}\tr\left(\bar\Y^\top\M\bar\Y-2\bar\Y^\top\W\bar\X\right)-\tr\left(\frac12\M^2-\W\C_{xx}\W^\top\right).
\end{align}
This step can be verified by differentiating $L(\W,\M,\bar\Y)$ with respect to $\W$ and $\M$ and noting that the optimal values are achieved when $\W$ and $\M$ equal $\frac{1}{T}\bar\Y\bar\X^\top\C_{xx}^{-1}$ and $\frac{1}{T}\bar\Y\bar\Y^\top$, respectively.
Finally, we interchange the order of minimization with respect to $\bar\Y$ and $\W$, as well as the order of optimization with respect to $\bar\Y$ and with respect to $\M$:
\begin{align}\label{eq:1b}
    &\min_{\W\in\R^{k\times m}}\max_{\M\in\mathcal{S}_{++}^k}\min_{\bar\Y\in\R^{k\times T}}L(\W,\M,\bar\Y).
\end{align}
The second interchange is justified by the fact that $L(\W,\M,\bar\Y)$ satisfies the saddle point property with respect to $\bar\Y$ and $\M$, which follows from the fact that $L(\W,\M,\bar\Y)$ is strictly convex in $\Y$ (since $\M$ is positive definite) and strictly concave in $\M$.

\subsection{Offline algorithm}

In the offline, or batch, setting, we have access to the sample covariance matrices $\C_{xx}$ and $\C_{\bar x\bar x}$, and we solve the min-max problem \eqref{eq:1b} by alternating optimization steps.
First, for fixed $\W$ and $\M$, we minimize the objective function $L(\W,\M,\bar\Y)$ over $\bar\Y$, to obtain
\begin{align}
    \label{eq:olZ}
    \bar\Y=\M^{-1}\W\bar\X.
\end{align}
With $\bar\Y$ fixed, we then perform a gradient descent-ascent step with respect to $\W$ and $\M$:
\begin{align}
    \label{eq:dW}
    \W&\gets\W+2\eta\left(\frac1T\bar\Y\bar\X^\top-\W\C_{xx}\right)\\
    \label{eq:dM}
    \M&\gets\M+\frac\eta\tau\left(\frac1T\bar\Y\bar\Y^\top-\M\right).
\end{align}
Here $\tau>0$ is the ratio of the learning rates of $\W$ and $\M$ and $\eta\in(0,\tau)$ is the (possibly time-dependent) learning rate for $\W$.
The condition $\eta<\tau$ ensures that matrix $\M$ remains positive definite given a positive definite initialization.

\subsection{Online algorithm}

In the online setting, the expanded signal $\{\x_t\}$ is streamed one sample at a time, and the algorithm must compute its output without storing any significant fraction of the data in memory.
In this case, at each time-step $t$, we compute the output $\y_t=\M^{-1}\a_t$, where $\a_t:=\W\x_t$ is the projection of $\x_t$ onto the $k$-dimensional ``slow'' subspace, in a biologically plausible manner by running the following fast (neural) dynamics to equilibrium (our algorithm implements these dynamics using an Euler approximation):
\begin{align}\label{eq:yode}
    \frac{d\y_t(\gamma)}{d\gamma}=\a_t-\M\y_t(\gamma).
\end{align}
To update the (synaptic) matrices $\W$ and $\M$, we replace the covariance matrices in \eqref{eq:dW}--\eqref{eq:dM} with the rank-1 stochastic approximations:
\begin{align*}
    \frac1T\bar\Y\bar\X^\top\mapsto\bar\y_t\bar\x_t^\top,&&\frac1T\bar\Y\bar\Y^\top\mapsto\bar\y_t\bar\y_t^\top,&&\C_{xx}\mapsto\x_t\x_t^\top.
\end{align*}
This yields the following stochastic gradient descent-ascent steps with respect to $\W$ and $\M$:
\begin{align*}
    \W&\gets\W+2\eta\left(\bar\y_t\bar\x_t^\top-\a_t\x_t^\top\right)\\
    \M&\gets\M+\frac{\eta}{\tau}\left(\bar\y_t\bar\y_t^\top-\M\right).
\end{align*}
We can now state our online SFA algorithm, which we refer to as Bio-SFA (Alg.~\ref{alg:online}). 

\begin{algorithm}[ht]
  \caption{Bio-SFA}
  \label{alg:online}
\begin{algorithmic}
  \STATE {\bfseries input} expanded signal $\{\x_0,\x_1,\dots,\x_T\}$; dimension $k$; parameters $\gamma$, $\eta$, $\tau$
  \STATE {\bfseries initialize} matrix $\W$ and positive definite matrix $\M$
  \FOR{$t=1,2,\dots,T$}
  \STATE $\a_t\gets\W\x_t$ \hfill $\triangleright\;$ project inputs
  \REPEAT
  \STATE $\y_t\gets\y_t+\gamma(\a_t-\M\y_t)$ \hfill $\triangleright\;$ compute neural output
  \UNTIL{convergence}
  \STATE $\bar\x_t\gets \x_t+\x_{t-1}$
  \STATE $\bar\y_t\gets\y_t+\y_{t-1}$ 
  \STATE $\W \gets \W + 2\eta (\bar\y_t\bar\x_t^\top - \a_t\x_t^\top)$ \hfill $\triangleright\;$ synaptic updates
  \STATE $\M \gets \M + \frac{\eta}{\tau} (\bar\y_t\bar\y_t^\top-\M) $
  \ENDFOR
\end{algorithmic}
\end{algorithm}

\section{Biologically plausible neural network implementation}
\label{sec:bio}

We now demonstrate that Bio-SFA can be implemented in a biologically plausible network, depicted in Fig.~\ref{fig:NN}.
Recall that we define a network to be biologically plausible if it computes its output in the online setting and has local learning rules.
The neural network consists of an input layer of $m$ neurons (blue circles) and an output layer of $k$ neurons with separate dendritic and somatic compartments (black circles with 2 compartments).
At each time $t$, the $m$-dimensional expanded signal $\x_t$, which is represented by the activity of the input neurons, is multiplied by the weight matrix $\W$, which is encoded by the feedforward synapses connecting the input neurons to the output neurons (green lines). 
This yields the $k$-dimensional projection $\a_t=\W\x_t$, which is represented in the dendritic compartment of the output neurons and then propagated to the somatic compartments.
This is followed by the fast recurrent neural dynamics Eq.~\eqref{eq:yode} amongst the somatic compartments of the output neurons, where the matrix $\M$ is encoded by the lateral synapses connecting the layer of output neurons (red lines).
These fast neural dynamics equilibriate at $\y_t=\M^{-1}\a_t$.
The $k$-dimensional output signal $\y_t$ is represented by the activity of the output neurons.

The synaptic updates are as follows.
Recall that $\bar\x_t=\x_t+\x_{t-1}$ (resp.\ $\bar\y_t=\y_t+\y_{t-1}$) is the delayed sum of the inputs (resp.\ outputs), which we assume are represented in the $m$ input neurons (resp.\ $k$ output neurons).
Biologically, they can be represented by slowly changing concentrations (e.g., calcium) at the pre- and post-synaptic terminals.
We can write the elementwise synaptic updates in Alg.~\ref{alg:online} as
\begin{align}
\label{eq:dWij}
    W_{ij}&\gets W_{ij}+2\eta\left(\bar y_t^i\bar x_t^j-a_t^ix_t^j\right), && 1\le i\le k,\;1\le j\le d,\\
    M_{ij}&\gets M_{ij}+\frac{\eta}{\tau}\left(\bar y_t^i\bar y_t^j-M_{ij}\right), && 1\le i,j\le k.
\end{align}
Since the $j$\textsuperscript{th} input neuron stores the variables $x_t^j,\bar x_t^j$ and the $i$\textsuperscript{th} output neuron stores the variables $a_t^i,y_t^i,\bar y_t^i$, the update for each synapse is local.

It is worth comparing the derived updates to the feedforward weights Eq.~\eqref{eq:dWij} to the updates proposed by F\"oldi\'ak~\cite{Fldik1991}, which are given by
\begin{align*}
    w_{ij}\gets w_{ij}+\eta\left(\bar y_t^i x_t^j-\bar y_t^i w_{ij}\right),&&1\le i\le k,\;1\le j\le d.
\end{align*}
The first terms in the updates, $\bar y_t^i\bar x_t^j$ and $\bar y_t^ix_t^j$, are quite similar.
The main difference between the updates is between the second terms: $a_t^ix_t^j$ and $\bar y_t^iw_{ij}$.
In our network, the second term $a_t^ix_t^j$ serves to whiten the inputs in our network, whereas F\"oldi\'ak's second term $\bar y_t^iw_{ij}$ is added as a decay to ensure the weights remain bounded.
In addition, our network includes lateral weights $M_{ij}$ which ensure that the projections $y_t^i$ are distinct, and such lateral weights are not included in F\"oldi\'ak's network.
While the updates are similar in some respects, it is difficult to compare the outputs of the networks because F\"oldi\'ak's network is postulated rather than derived from a principled objective function, so the network must be simulated numerically in order to evaluate its output.

\section{Experiments}
\label{sec:numerics}

To validate our approach, we test Bio-SFA (Alg.~\ref{alg:online}) on synthetic and naturalistic datasets.
We provide an overview of the experiments here and defer detailed descriptions and additional figures to Sec.~\ref{apdx:numerics} of the supplement.
The evaluation code is available at \url{github.com/flatiron/bio-sfa}.

To measure the performance of our algorithm, we compare the ``slowness'' of the projection $\Y=\M^{-1}\W\X$, with the slowest possible projection.
This can be quantified using the objective \eqref{eq:min}.
We first evaluate the objective \eqref{eq:min} at its optimum:
\begin{align*}
    \lambda_{\text{slow}}:=\min\left\{\tr\V^\top\C_{\dot x\dot x}\V:\V\in\R^{m\times k}\;\text{s.t.}\;\V^\top\C_{xx}\V=\I_k\right\}
\end{align*}
which can be evaluated using an offline generalized eigenvalue problem solver.
To compute the error at each iteration, we compare the slowness of the current projection to the minimal slowness:
\begin{align}
    \label{eq:error}
    \text{Error}=\tilde\V^\top\C_{\dot x\dot x}\tilde\V-\lambda_{\text{slow}},\qquad\tilde\V:=\W^\top\M^{-1}(\M^{-1}\W\C_{xx}\W^\top\M^{-1})^{-1/2},
\end{align}
where the normalization ensures that $\tilde\V$ satisfies the constraint in Eq.~\eqref{eq:min}.
In Sec.~\ref{apdx:numerics}, we show that $\V$ indeed asymptotically satisfies the constraint in Eq.~\eqref{eq:min}.

\subsection{Chaotic time series}

\begin{figure}[ht]
\centering
\subfloat[Driving force]{\label{fig:driving}\includegraphics[height=0.24\textwidth]{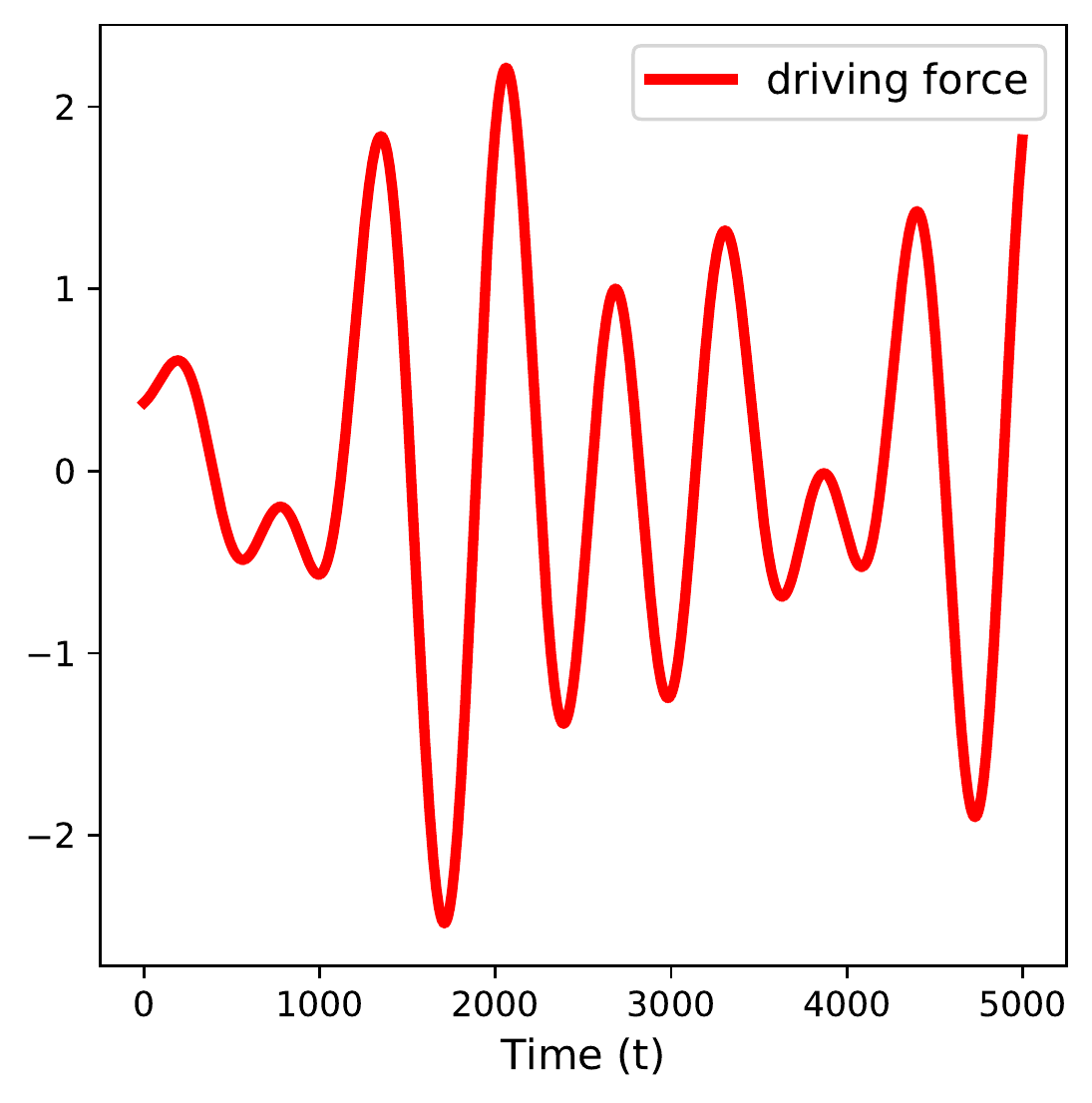}}
\hfill
\subfloat[Noisy series]{\label{fig:raw}\includegraphics[height=0.24\textwidth]{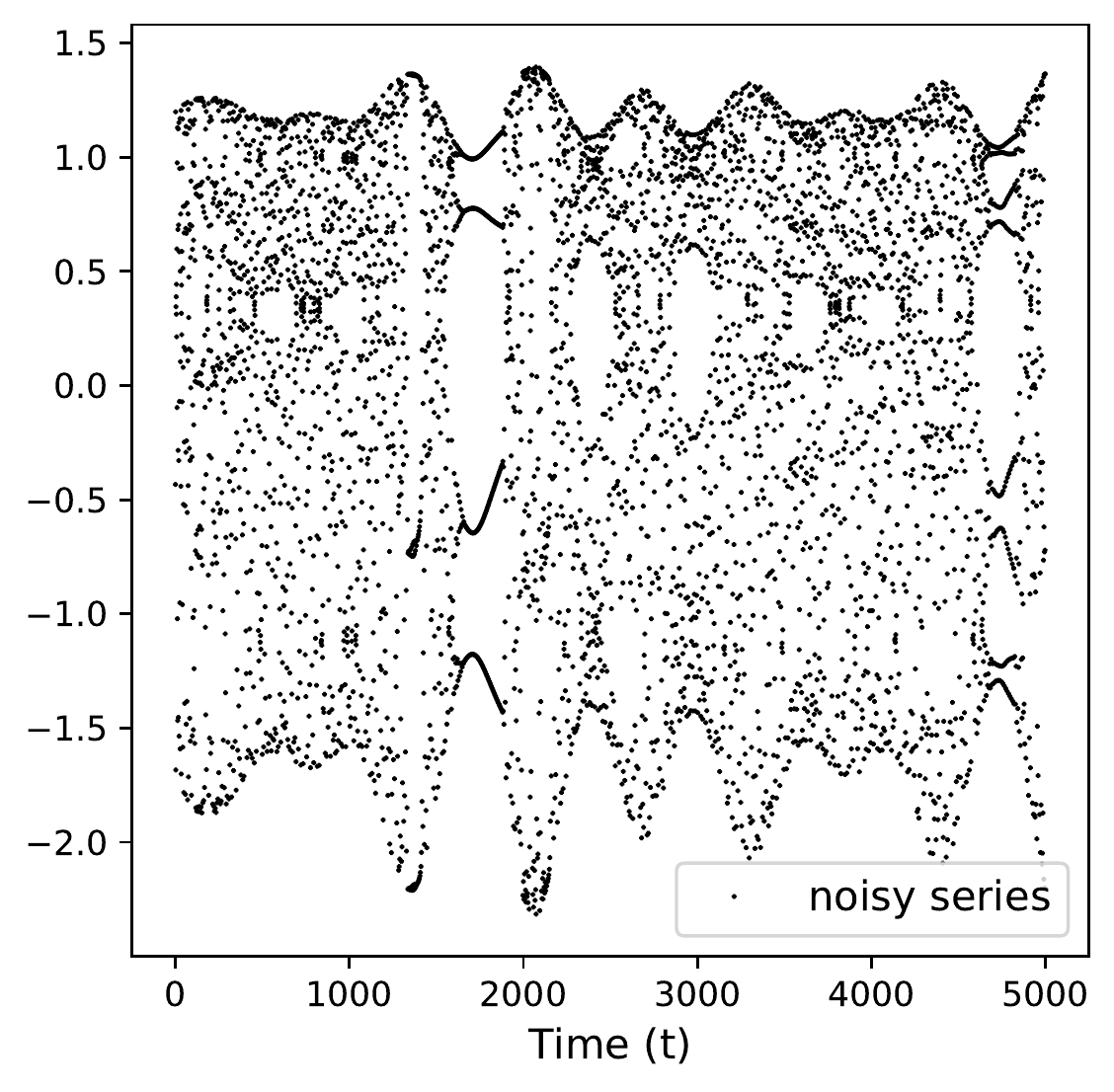}}
\hfill
\subfloat[Bio-SFA output]{\label{fig:slow}\includegraphics[height=0.24\textwidth]{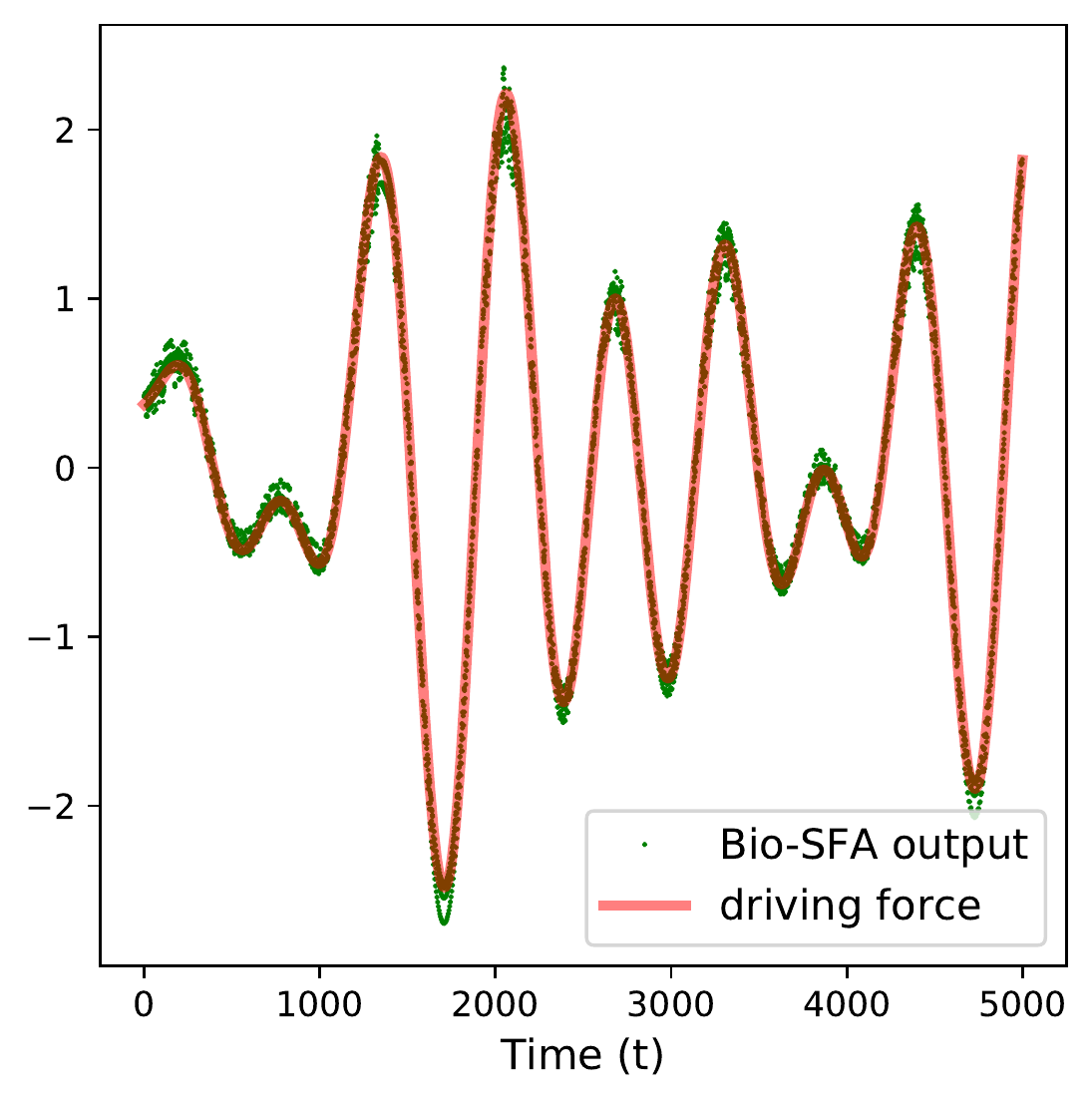}}
\hfill
\subfloat[Error]{\label{fig:slow_obj}\includegraphics[height=0.24\textwidth]{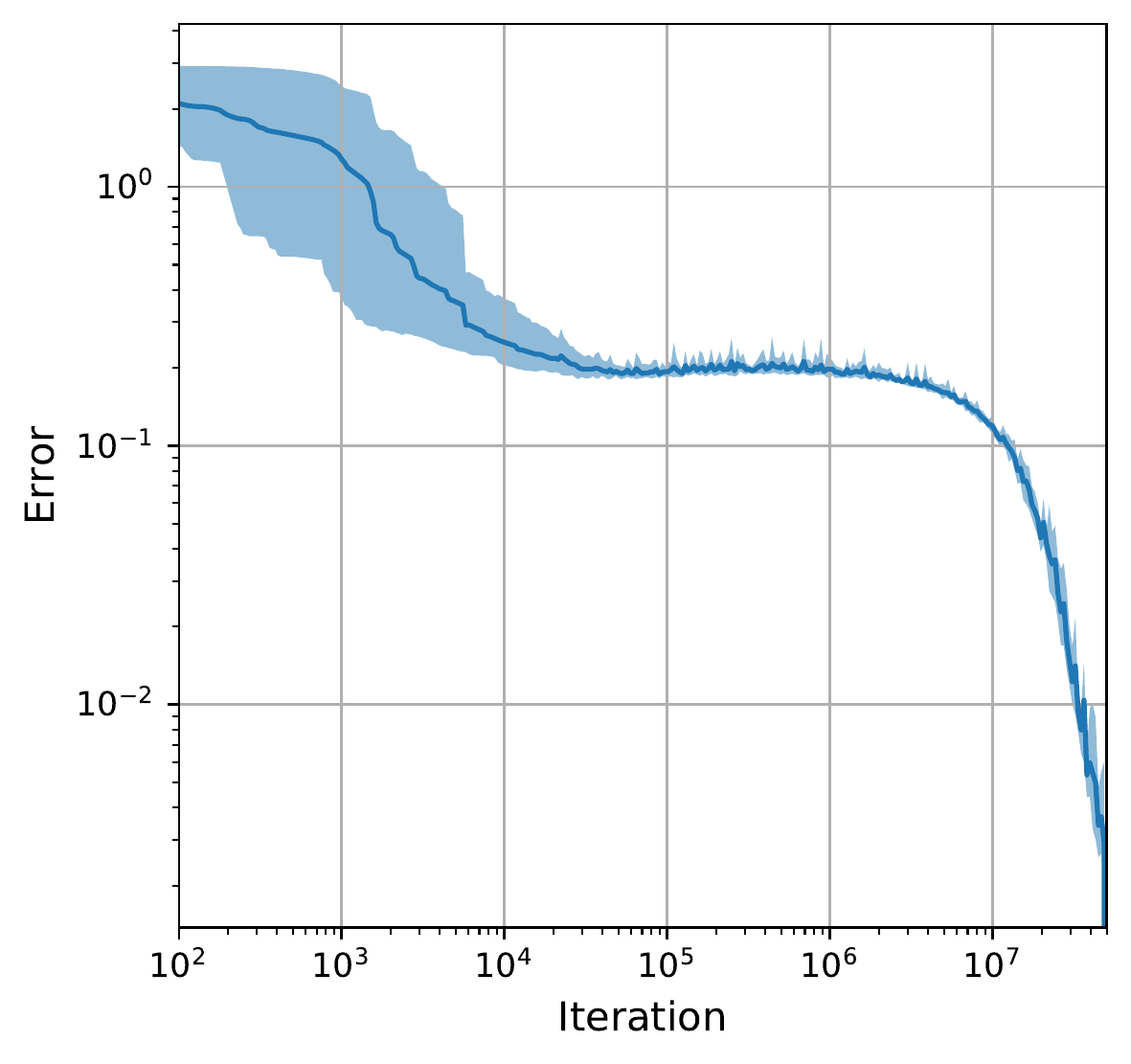}}
\caption{Performance of Bio-SFA on a noisy series generated by a logistic map with slow driving force. Panels (a), (b) and (c) depict the final 5000 time steps (out of $5\times 10^7$ time steps) of the normalized driving force $\{\gamma_t\}$ (red line), noisy series $\{z_t\}$ (black dots), and Bio-SFA output $\{y_t\}$ (green dots). Panel (d) shows the mean error and 90\% confidence intervals over 10 runs.}
\end{figure}

Before testing on naturalistic datasets, we test Bio-SFA on a challenging synthetic dataset.
Let $\{\gamma_t\}$ be a (slow) driving force equal to the sum of 6 sine functions with random amplitudes, frequencies and phases, Fig.~\ref{fig:driving} (red line).
Define the noisy series derived from the recursive logistic map with time-varying growth rate: $z_t=(3.6 + 0.4\gamma_t)z_{t-1}(1-z_{t-1})$, Fig.~\ref{fig:raw} (black dots).
Wiskott \cite{wiskott2003estimating} showed that the driving force $\{\gamma_t\}$ can be recovered from the noisy series $\{z_t\}$ by implementing (offline) Quadratic SFA on the 4-dimensional signal $\{\s_t\}$ whose components correspond to the values of the noisy series over the 4 most recent time steps, i.e., $\s_t:=(z_t,z_{t-1},z_{t-2},z_{t-3})$.
We replicate the results from \cite{wiskott2003estimating} using Bio-SFA.
Let $\{\x_t\}$ be the 14-dimensional quadratic expansion of $\{\s_t\}$.
We use Bio-SFA to extract the slowest one-dimensional projection $\{y_t\}$, Fig.~\ref{fig:slow} (green dots).
Qualitatively, we see that the slowest projection recovered by Bio-SFA closely aligns with the slow driving force $\{\gamma_t\}$.
In Fig.~\ref{fig:slow_obj} we plot the error at each iteration.

\subsection{Sequence of natural images}

\begin{figure}[ht]
\centering
\subfloat[Image sequence generation]{\label{fig:offlinegabors}\includegraphics[height=0.306\textwidth]{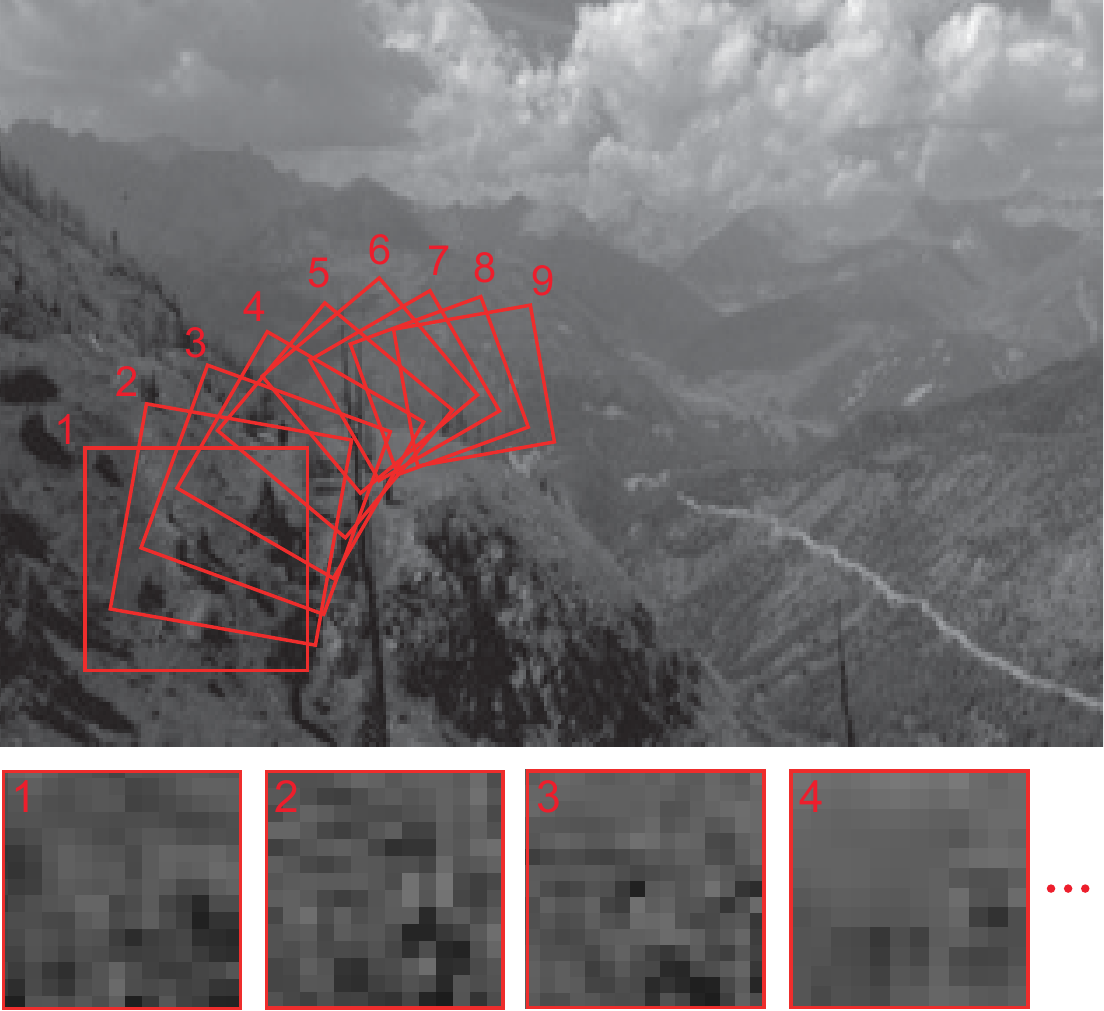}}
\hfill
\subfloat[Optimal stimuli]{\label{fig:onlinegabors}\includegraphics[height=0.31\textwidth]{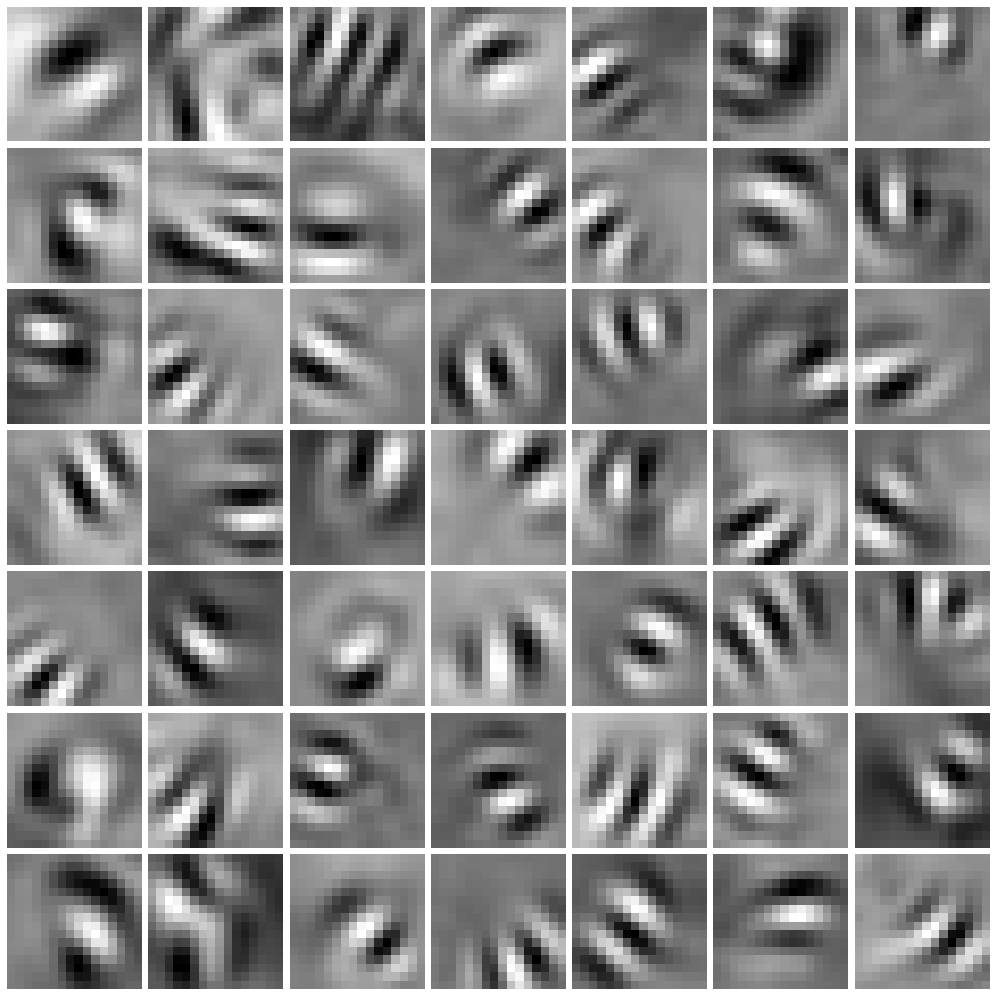}}
\hfill
\subfloat[Error]{\label{fig:error}\includegraphics[height=0.314\textwidth]{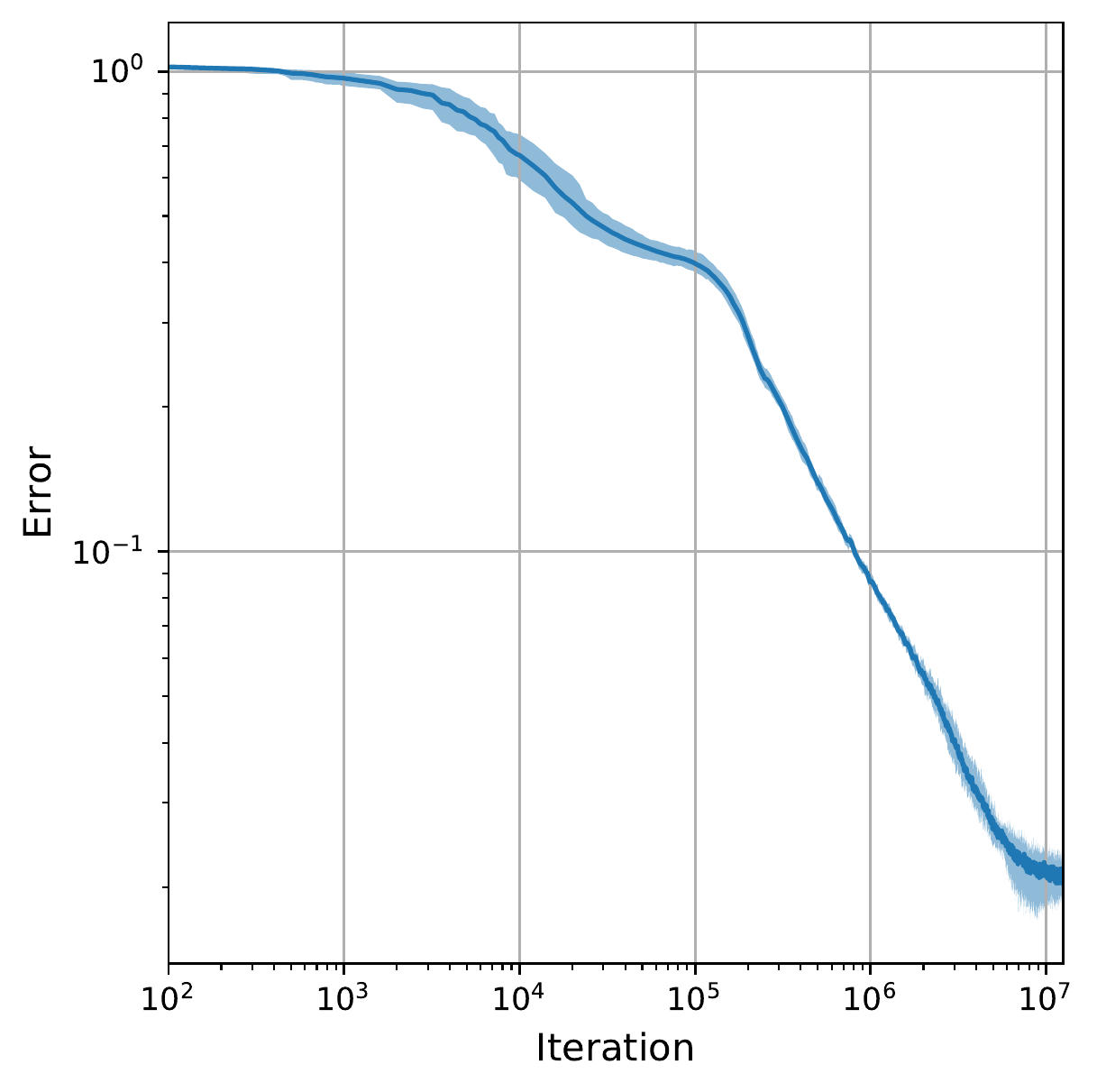}}
\caption{Performance of Bio-SFA on a sequence of natural images. Panel (a) illustrates the generation of the sequence. Panel (b) shows the maximally excitatory stimuli for the 49-dimensional output obtained by Bio-SFA. 
Panel (c) depicts the mean error and 90\% confidence intervals over 10 runs.}
\label{fig:filters}
\end{figure}

Next, we test Bio-SFA on a sequence of natural images.
First, a 256-dimensional sequence $\{\z_t\}$ was generated by moving a $16\times 16$ patch over 13 natural images from \cite{hyvarinen2000independent} via translations, zooms, and rotations, Fig.~\ref{fig:offlinegabors}.
To extract relevant features, we follow the exact same procedure as Berkes and Wiskott \cite{berkes2002applying}, but replace the offline SFA solver with Bio-SFA to generate a 49-dimensional output signal $\{\y_t\}$.
To visualize the 49-dimensional output, we calculate the unit vector $\z\in\R^{256}$ that maximizes $y^i$, for $i=1,\dots,49$.
These optimal stimuli, $\z$, which are displayed as $16\times 16$ patches in Fig.~\ref{fig:onlinegabors}, resemble Gabor patches and are in qualitative agreement with physiological characteristics of complex cells in the visual cortex.
This aligns with the results in \cite{berkes2002applying}; see also, \cite{berkes2005slow}.
To evaluate the performance of Bio-SFA, we plot the error at each iteration in Fig.~\ref{fig:error}.

\subsection{Hierarchical SFA on the visual stream of a simulated rat}

\begin{figure}[ht]
\centering
\subfloat[Layered architecture]{\label{fig:layers}\includegraphics[height=0.24\textwidth]{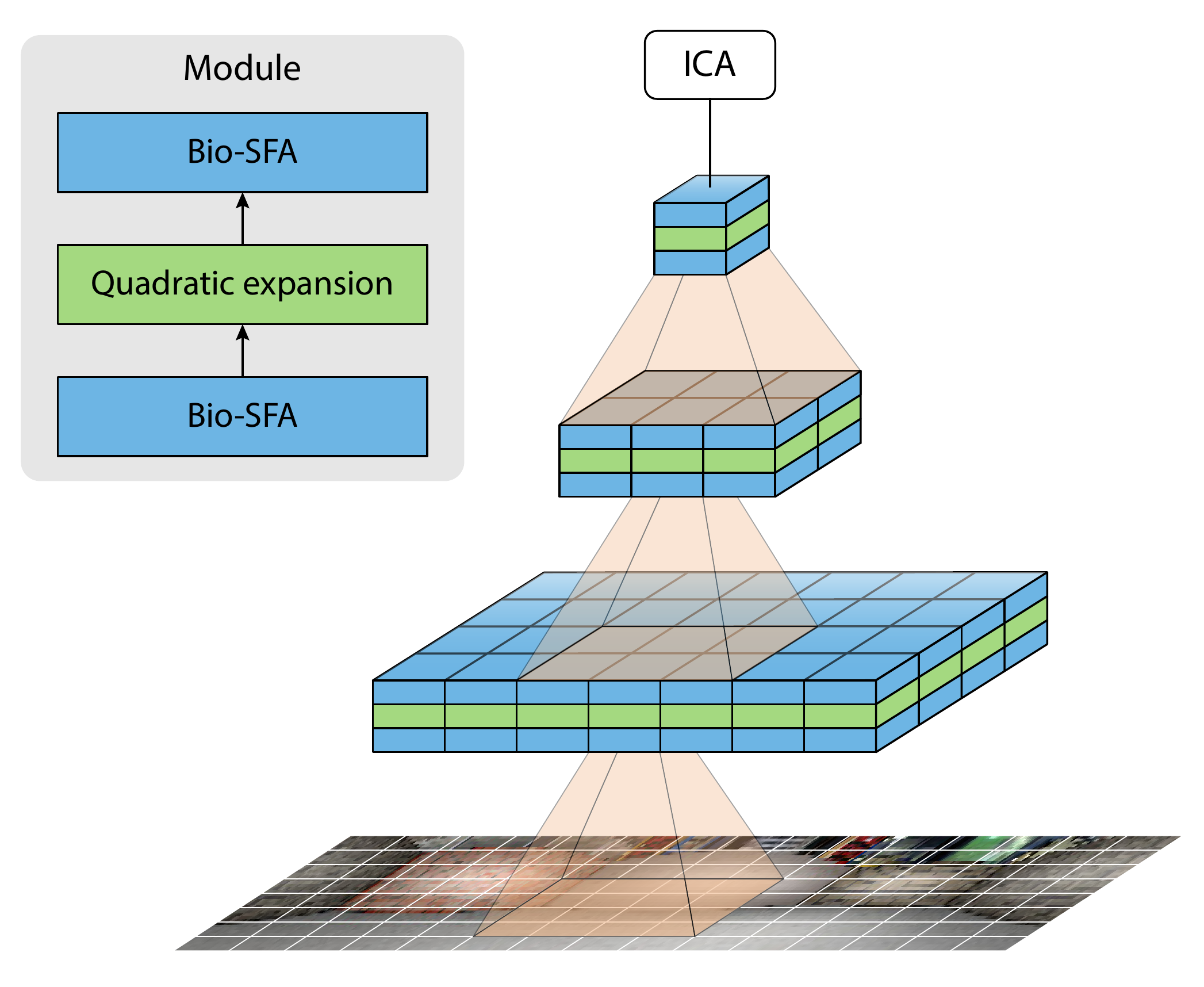}}
\hfill
\subfloat[SFA firing maps]{\label{fig:hsfa_firing}\includegraphics[height=0.24\textwidth]{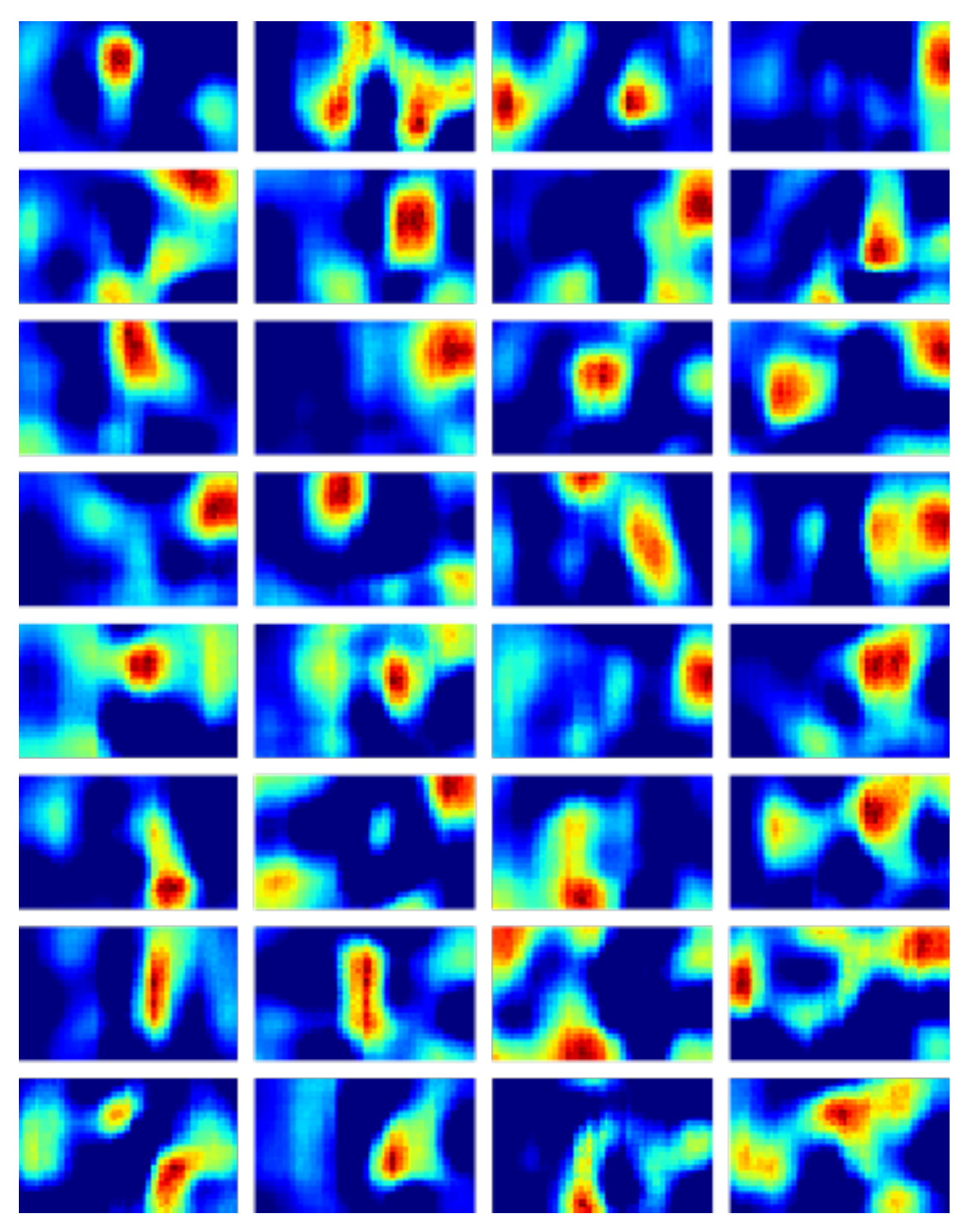}}
\hfill
\subfloat[ICA firing maps]{\label{fig:ica_firing}\includegraphics[height=0.24\textwidth]{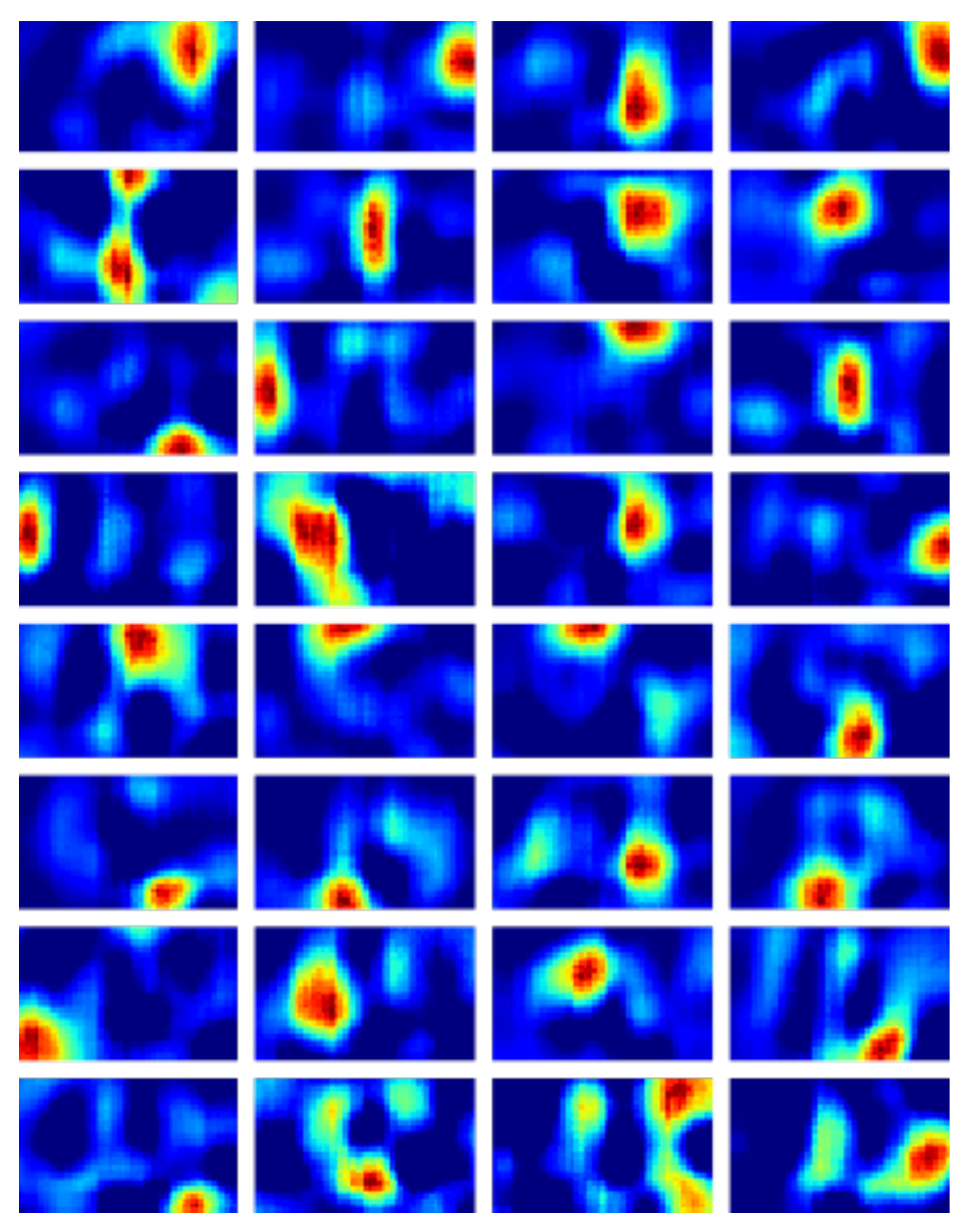}}
\hfill
\subfloat[Slowness of SFA output]{\label{fig:hsfa_error}\includegraphics[height=0.24\textwidth]{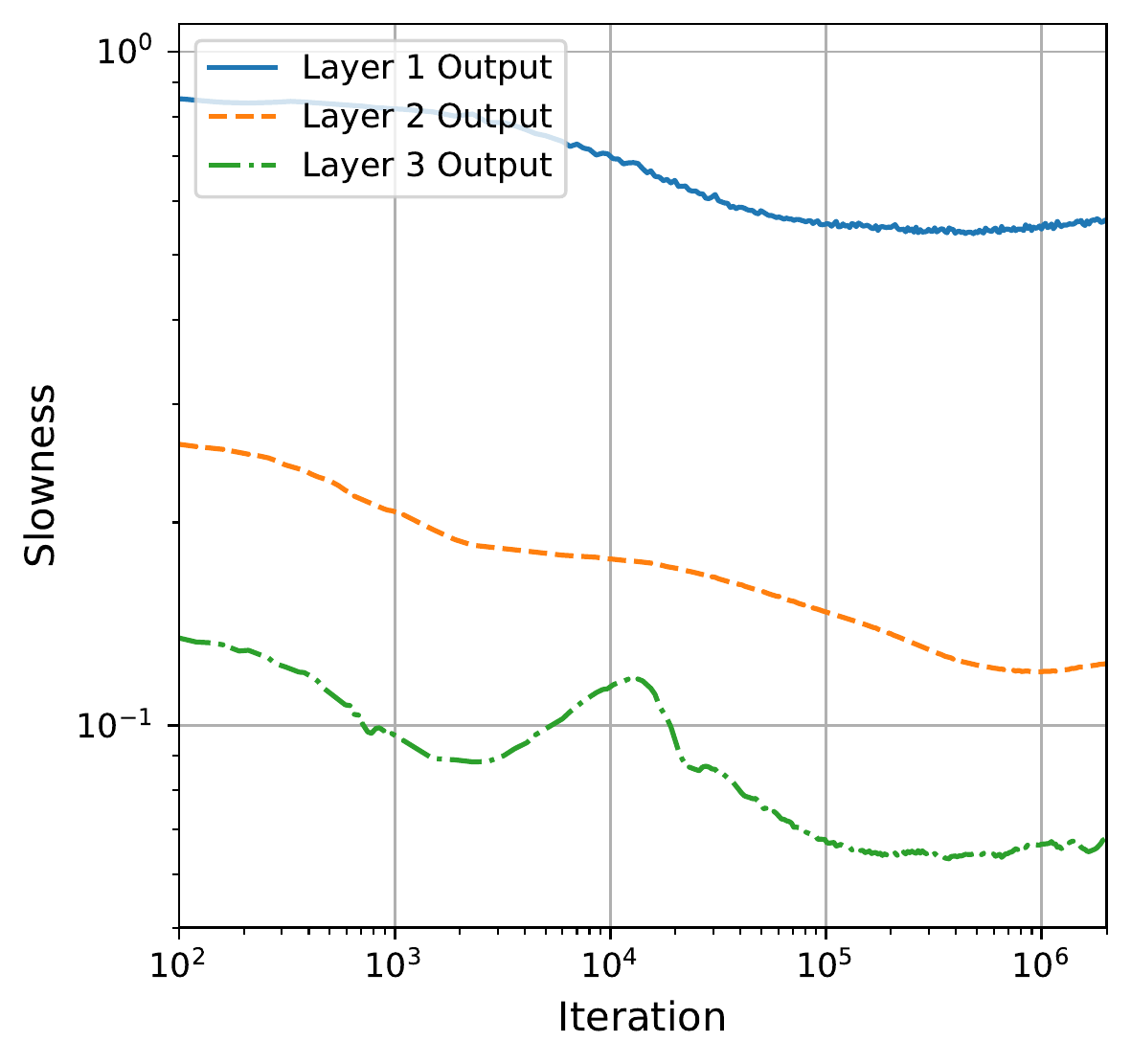}}
\caption{Performance of hierarchical Bio-SFA on a visual stream of a simulated rat. Panel (a) displays a schematic of the layered architecture and the operations within each module. Panels (b) and (c) depict the firing maps of the units in the final SFA layer and the subsequent ICA layer. Each rectangle shows the response of a component of the output as a function of the simulated rat's position within the rectangular room. Panel (d) shows the slowness of each layer's output at each iteration for a single trial.}
\label{fig:firingmaps}
\end{figure}

Following Sch\"onfeld and Wiskott \cite{schonfeld2015hsfa}, we test a hierarchical 3-layer organization of Bio-SFA ``modules'' on the inputs from the RatLab framework \cite{schonfeld2013ratlab}, which simulates the field of view of a rat with random trajectories in a rectangular room.
Each layer consists of spatially distributed modules that receive overlapping patches of either the visual stream or the preceding layer.
Inside each module, there are 3 steps: (1) Bio-SFA first reduces the dimension of the inputs to generate a 32-dimensional signal, (2) the reduced signal is quadratically expanded, and (3) Bio-SFA reduces the expanded signal to the slowest 32 features.
The layers are organized so that the modules in each successive layer receive inputs from larger patches of the visual field, Fig.~\ref{fig:layers}.
Adopting the procedure in \cite{schonfeld2015hsfa}, the network is trained greedily layer-by-layer with weight sharing across modules in each layer (see Sec.~\ref{apdx:numerics} of the supplement).
The final layer consists of a single module, with a 32-dimensional output, whose spatially-dependent firing maps are shown in Fig.~\ref{fig:hsfa_firing}.
The 3 SFA layers are followed by a fourth layer, which performs sparse coding via Independent Component Analysis (ICA) \cite{hyvarinen1999fast} (in the offline setting) with a 32-dimensional output, whose firing map is shown in Fig.~\ref{fig:ica_firing}.
As in \cite{schonfeld2015hsfa}, the firing maps of the final ICA layer are spatially localized and resemble the firing maps of place cells in the hippocampus.
To quantify the performance of this hierarchical network, we plot the slowness (not errors, see Sec.~\ref{apdx:numerics} of the supplement) of each of the first 3 layers' outputs at each iteration, Fig.~\ref{fig:hsfa_error}.

\section{Discussion}

We derived an online algorithm for SFA with a biologically plausible neural network implementation, which is an important step towards understanding how the brain could use temporal slowness as a computational principle.
While our network implementation satisfies natural requirements for biological plausibility, it differs from biological neural circuits in a number of ways.
For instance, our network includes direct lateral inhibitory synapses between excitatory neurons, whereas inhibition is typically modulated by interneurons in biological networks.
By adapting the approach in \cite{pehlevan2015normative}, interneurons can be introduced to modulate inhibition.
Second, the synaptic updates in our network require both the pre- and post-synaptic neurons to store slow variables; however, signal frequencies in dendrites are slower than in axons, suggesting that it is more likely for slow variables to be stored in the post-synaptic neuron, not the pre-synaptic neuron.
We can address this with a modification, which is exact when the expanded signal $\{\x_t\}$ exhibits time-reversibility, so that only the post-synaptic represents slow variables; see Sec.~\ref{sec:rev} of the supplement.
Finally, our network includes linear neurons, which do not respect the nonnegativity constraints of neuronal outputs.
An interesting future direction is to understand the effect of enforcing a nonnegativity constraint on $\y_t$ in the objective function \eqref{eq:similarity}.

\section*{Broader impact}

An important problem in neuroscience is to understand the computational principles the brain uses to process information.
Progress on this front has the potential to have wide ranging benefits for helping to manage the adverse effects of neurological diseases and disorders.
This work represents a small step in that direction.

\section*{Acknowledgements}

We thank Yanis Bahroun, Nicholas Chua, Shiva Farashahi, Johannes Friedrich, Alexander Genkin, Jason Moore, Anirvan Sengupta and Tiberiu Tesileanu for helpful comments and feedback on an earlier draft of this work.

\bibliography{sfa.bib}
\bibliographystyle{plain}

\clearpage

\appendix

\section*{Supplemental material}


\section{Detailed derivation of the SFA objective}
\label{apdx:derivation}

Here, we provided a detailed derivation of the SFA objective \eqref{eq:similarity}.
We allow for the case that $\C_{xx}$ is not full rank (but is at least rank $k$).

Our starting point is the objective in Eq.~\eqref{eq:min}, which we recall here:
\begin{align}
\label{eq:min_apdx}
    \argmin{\V\in\R^{m\times k}}\tr\V^\top\C_{\dot x\dot x}\V\quad\text{subject to}\quad\V^\top\C_{xx}\V=\I_k.
\end{align}
Under the whitening constraint $\V^\top\C_{xx}\V=\I_k$, we have the following relation: 
\begin{align*}
    \V^\top\C_{\dot x\dot x}\V=2\I_k-\frac1T\sum_{t=1}^T\V^\top(\x_t\x_{t-1}^\top+\x_{t-1}\x_t^\top)\V=4\I_k-\V^\top\C_{\bar x\bar x}\V.
\end{align*}
Since $\tr\I_k=k$ is constant, it does not affect the output of the argmin.
Therefore, we can rewrite the objective in Eq.~\eqref{eq:min_apdx} as the following maximization problem:
\begin{align}
\label{eq:max_apdx}
    \argmax{\V\in\R^{m\times k}}\tr\V^\top\C_{\bar x\bar x}\V\quad\text{subject to}\quad\V^\top\C_{xx}\V=\I_k.
\end{align}
Next, let $k\le n\le m$ be the rank of $\C_{xx}$. 
We first project $\X$ onto its $n$-dimensional principal subspace; i.e., onto the subspace spanned by eigenvectors of $\C_{xx}$ corresponding to positive eigenvalues.
To this end, consider the eigendecomposition $\C_{xx}=\U\Lam\U^\top$, where $\Lam$ is an $n\times n$ diagonal matrix whose diagonal elements are the positive eigenvalues of $\C_{xx}$ and $\U$ is a $m\times n$ matrix whose orthonormal column vectors are the corresponding eigenvectors.
Then $\U\U^\top\in\R^{m\times m}$ is the matrix that projects $\X$ onto its $n$-dimensional principal subspace and Eq.~\eqref{eq:max_apdx} is equivalent to the maximization problem:
\begin{align}
\label{eq:maxU_apdx}
    \argmax{\V\in\R^{m\times k}}\tr\V^\top\U\U^\top\C_{\bar x\bar x}\U\U^\top\V\quad\text{subject to}\quad\V^\top\U\U^\top\C_{xx}\U\U^\top\V=\I_k.
\end{align}
Setting $\hat\x_t:=\Lam^{-1/2}\U^\top\bar\x_t$ for $t=1,\dots,T$, and
\begin{align*}
    \hat\V:=\Lam^{1/2}\U^\top\V,&&\C_{\hat x\hat x}:=\frac1T\sum_{t=1}^T\hat\x_t\hat\x_t^\top=\Lam^{-1/2}\U^\top\C_{\bar x\bar x}\U\Lam^{-1/2},
\end{align*}
we see that $\hat\V$ is the solution of:
\begin{align}
    \label{eq:pca_apdx}
    \argmax{\hat\V\in\R^{m\times k}}\tr\hat\V^\top\C_{\hat x\hat x}\hat\V\quad\text{subject to}\quad\hat\V^\top\hat\V=\I_k.
\end{align}
Eq.~\eqref{eq:pca_apdx} is the variance maximization objective for the PCA eigenproblem, which is optimized when the column vectors of $\hat \V$ span the $k$-dimensional principal subspace of $\C_{\hat x\hat x}$.

Finally, define the data matrices
\begin{align*}
    \bar\X:=[\bar\x_t,\dots,\bar\x_T],&&\hat\X:=[\hat\x_1,\dots,\hat\x_T],&&\bar\Y:=[\bar\y_1,\dots,\bar\y_T].
\end{align*}
Then, since $\bar\y_t=\V^\top\U\U^\top\bar\x_t=\hat\V^\top\hat\x_t$, we see that $\bar\Y$ is the projection of $\hat\X_t$ onto its $k$-dimensional principal subspace.
As shown in \cite{cox2000multidimensional}, this principal projection can be expressed as a solution of the following objective from classical multidimensional scaling:
\begin{align*}
  \argmin{\bar\Y\in\R^{k\times T}}\frac{1}{2T^2}\big\|\bar\Y^\top\bar\Y-\hat\X^\top\hat\X\big\|_\text{Frob}^2=\argmin{\bar\Y\in\R^{k\times T}}\frac{1}{2T^2}\big\|\bar\Y^\top\bar\Y-\bar\X^\top\C_{xx}^+\bar\X\big\|_\text{Frob}^2,
\end{align*}
where we have used the fact that $\hat\X^\top\hat\X=\bar\X^\top\U\Lam^{-1}\U^\top\bar\X=\bar\X^\top\C_{xx}^+\bar\X$.
Lastly, we note that the minimization problem in Eq.~\eqref{eq:minimization} is equivalent to the min-max problem in Eq.~\eqref{eq:L} with $\C_{xx}^+$ in place of $\C_{xx}^{-1}$.
This can be verified by differentiating $L(\W,\M,\bar\Y)$ with respect to $\W$ and noting that the optimal value is achieved when $\W$ equals $\frac{1}{T}\bar\Y\bar\X^\top\C_{xx}^{+}$.

\clearpage

\section{Experimental methods}
\label{apdx:numerics}

In this section, we detail how we implemented each experiment.

\subsection{Implementation of neural dynamics}

In Bio-SFA, to compute the output $\y_t=\M^{-1}\a_t$, we use the recursive updates $\y_t\gets\y_t+\gamma(\a_t-\M\y_t)$ because they respect the neural architecture.
For purposes of simulation, we multiply $\M^{-1}$ by $\a_t$ to compute the output.
When $k>1$, to speed up simulations, we store the value of $\M^{-1}$ and make rank-1 updates at each iteration using the Sherman-Morrison formula.

\subsection{Chaotic time series}

\paragraph{Driving force:} The driving force $\{\gamma_t\}$ is defined to be the sum of 6 sine functions, as follows:
\begin{align*}
    \gamma_t:=\sum_{i=1}^6A_i\sin(\theta_i t+\omega_i),\qquad t=1,2,\dots.
\end{align*}
Here the amplitudes $A_1,\dots,A_6$ are uniformly sampled from the interval $(.1,2)$ and then normalized so that they sum to 1, the frequencies $\theta_1,\dots,\theta_6$ are uniformly sampled from the interval $(0.25,1.25)$, and the phases $\omega_1,\dots,\omega_6$ are uniformly sampled from the interval $(0,2\pi)$.

\paragraph{Hyperparameters:} We used the learning rate $\eta_t=1/(a+bt)$.
To choose our hyperparameters, we performed a grid search over $a\in\{10^2,10^3,10^4,10^5\}$, $b\in\{10^{-1},10^{-2},10^{-3},10^{-4}\}$ and $\tau\in\{0.01,0.05,0.1,0.5,1,5\}$.
We found the optimal hyperparameters to be $a=4$, $b=-4$ and $\tau=0.5$.

\paragraph{Hardware:} The experiment was performed on a 2.8 GHz Quad-Core Intel Core i7 CPU.

\subsection{Sequence of natural images}

\paragraph{Implementation:}

We extracted 2,500 image sequences from the 13 images used in \cite{hyvarinen2000independent}. 
To generate each sequence, one of the 13 images was chosen uniformly at random, and then a sequence of 100 $16\times 16$ patches was extracted following the procedure in \cite{berkes2002applying,berkes2005slow}, using the default parameters from the code released with those papers. 
In particular, sums of sinusoids with random phases and amplitudes are used to drive the translation, zoom and rotation of the $16\times 16$ field of view. 
Following \cite{berkes2002applying,berkes2005slow}, we first project the 254-dimensional image sequence onto its 64-dimensional principal subspace.
Bio-SFA is then trained on the 2144-dimensional quadratic expansion of the 64-dimensional projected sequence.

\paragraph{Hyperparameters:}
We used the learning rate $\eta_t=\alpha/(1+t/\beta)$.
To choose our hyperparameters, we performed a grid search over $\alpha\in\{2.5\times10^{-6},5\times10^{-6},2.5\times10^{-5},5\times10^{-5},2.5\times10^{-4},5\times10^{-4}\}$, $\beta\in\{10^4,10^6,5\times10^6,10^7,5\times10^7,10^8,10^9,10^{10}\}$ and $\tau\in\{0.5,1,2,4\}$.
We found the optimal hyperparameters to be $\alpha=5\times10^{-6}$, $\beta=5\times10^6$ and $\tau=1$.

\paragraph{Orthonormality constraint:}

\begin{figure}[ht]
    \centering
    \includegraphics[width=0.5\textwidth]{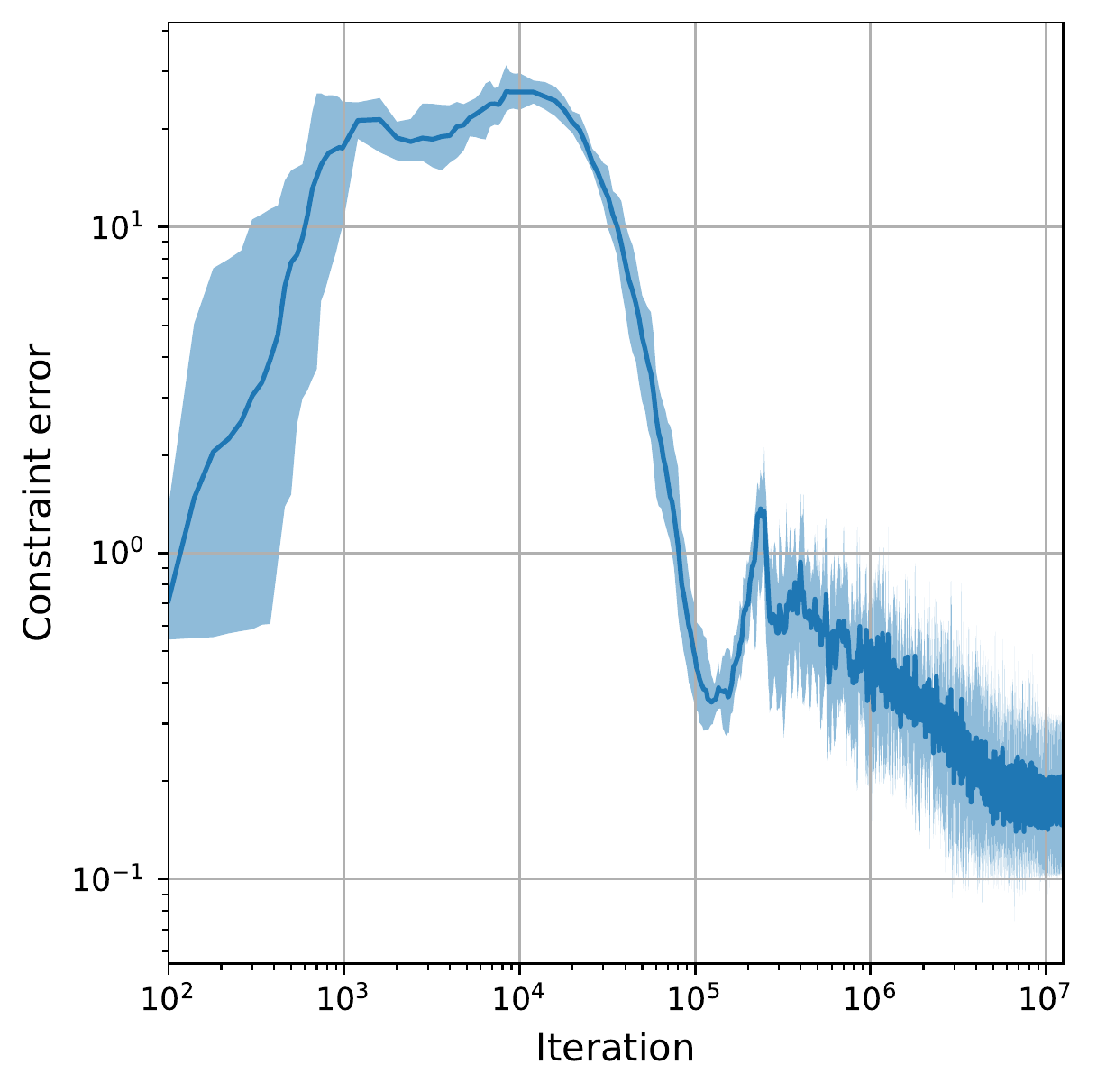}
    \caption{Convergence of the constraint error defined in Eq.~\eqref{eq:constraint_error_supp} for Bio-SFA (Alg.~\ref{alg:online}). The lines and shaded regions show the mean error and 90\% confidence intervals over ten runs.}
    \label{fig:constraint_supp}
\end{figure}

To evaluate if Bio-SFA satisfies the orthonormality constraint $\V^\top\C_{xx}\V=\I_k$ in Eq.~\eqref{eq:genEV}, where $\V=\W^\top\M^{-1}$, we use the normalized squared Frobenius norm, defined as follows:
\begin{equation}
    \label{eq:constraint_error_supp}
    \text{Constraint error}=\frac{1}{k}\,\left\| \M^{-1}\W\C_{xx}\W^\top\M^{-1} - \I_k\right\|^2_{\text{Frob}}.
\end{equation}
In Fig.~\ref{fig:constraint_supp}, we plot the constraint error at each iteration.

\paragraph{Hardware:} The experiment was performed on an NVIDIA Tesla V100 GPU.

\subsection{Hierarchical SFA on the visual stream of a simulated rat}

\paragraph{Simulated visual stream:}
To generate the input data to the hierarchical network, a sequence of 10,000 samples from the default scene in RatLab \cite{schonfeld2013ratlab} was generated, following the open field experiments in \cite{schonfeld2015hsfa}. RatLab simulates a rat's motion by driving its linear and angular momentum by random signals chosen to match experimental data. A wide image is extracted to match the rat's wide field of view. The resulting image sequence is used directly as training data for the online experiments here, after centering and rescaling.

\paragraph{Architecture:}

The hierarchical organization consists of 3 layers of Bio-SFA ``modules'', described below, followed by a fourth ICA layer; see Fig.~\ref{fig:layers}.
The input to the layered architecture is a sequence of $320\times 40$ color images.
The output of all SFA layers and the ICA layer are sequences of 32-dimensional vectors.

\paragraph{Description of the layers:}
The 4 layers are as follows:
\begin{itemize}
    \item[1.]The first layer consists of a 2-dimensional array of $63\times 9$ Bio-SFA modules.
    Each module receives as input $10\times 8$ pixel patches sampled from the $320\times40$ input, with each patch offset from its neighbors by half of the receptive field width in each dimension.
    The patches are then transformed into $240=80\times3$-dimensional vectors to be passed into the modules.
    The output of each module is a sequence of 32-dimensional vectors.
    \item[2.] The second layer consists of a 2-dimensional array of $8\times 2$ Bio-SFA modules.
    Each module receives inputs from a $14\times 6$ grid of modules from the first layer, again overlapping each other by half their length in each dimension.
    Since the output of each module in the first layer is 32-dimensional, the vectorized input to each module has dimension $2688=32\times 14\times 6$.
    The output of each module in the second layer is a sequence of 32-dimensional vectors.
    \item[3.] The third layer consists of a single Bio-SFA module that receives input from all $8\times 2$ modules in the second layer.
    Thus, the input to the third layer module has dimension $512=32\times 8\times 2$.
    The output of the third layer is a sequence of 32-dimensional vectors.
    \item[4.] The fourth layer is an offline ICA algorithm, described below.
    It receives as input the 32-dimensional vector output of the third layer and produces a 32-dimensional output.
\end{itemize}

\paragraph{Description of a Bio-SFA module:}
Each module receives a sequence of vector inputs (whose dimension depends on the layer) and outputs a 32-dimensional sequence.
The module consists of 3 steps:
\begin{itemize}
    \item[1.] Bio-SFA is applied to the input sequence to generate the slowest 32-dimensional projection.
    \item[2.] The projected sequence is quadratically expanded to generate the 560-dimensional expanded sequence, which is centered in the online setting using the running mean.
    \item[3.] Bio-SFA is applied to the expanded sequence to generate a 32-dimensional output.
\end{itemize}

\paragraph{Description of the ICA layer:} After an online hierarchy is trained, it is exported to a hierarchy of MDP (Modular toolkit for Data Processing) \cite{Zito2008mdp} nodes that can be read by the RatLab framework \cite{schonfeld2013ratlab}. Then, RatLab is instructed to fit an ICA layer in the offline setting, using MDP's implementation of CuBICA \cite{blaschke2004cubica}.

\paragraph{Training procedure:} Following \cite{franzius2007hsfa}, the layers were trained in a greedy layer-wise fashion, i.e., the layers are trained sequentially and the weights in a layer are fixed once it has been trained.
The Bio-SFA layers are trained using weight sharing; that is, each layer uses the same synaptic weights $\W$ and $\M$, which are shared across all patches.
To compute the $\W$ and $\M$ updates at each training step, the updates for each patch is computed according to Alg.~\ref{alg:online}, and these updates are summed to generate the updates for $\W$ and $\M$, which are scaled by the square root of the number of patches.

We use time-dependent learning rates of the form $\eta_t=\alpha/(1+t/\beta)$, with $\beta$ fixed to $5\times10^6$ in all modules.
For the first Bio-SFA step in the first module, we set $\alpha=5\times10^{-7}$ and $\tau=5\times10^{-4}$. In the rest of the modules, the first Bio-SFA step used $\alpha=2.5\times 10^{-6}$ with the same $\tau$.
For the second Bio-SFA step in each module, we set $\alpha=5\times10^{-5}$ and $\tau=1$.

\paragraph{Firing maps:} To generate a firing map from either the final SFA layer or the ICA layer, RatLab is instructed to generate a test set of still images by sampling the visual field of the simulated rat across a fine grid of spatial positions, using 8 head directions at each location. The output activities from each of the 32 units in either the final SFA or ICA layer are averaged over head orientation to generate a heatmap of that unit's activity over the spatial grid. Those maps are shown in Fig.~\ref{fig:firingmaps}.

\paragraph{Quantification of slowness:} To demonstrate that each layer is finding slower features, we plot the ``slowness'' of each layer's output, which is defined by
\begin{align*}
    \text{Slowness}=\tilde\V^\top\C_{\dot x\dot x}\tilde\V,
\end{align*}
where $\tilde\V$ is defined as in Eq.~\eqref{eq:error} and $\C_{\dot x\dot x}$ denotes the covariance of the discrete-time derivative of the expanded input for that module.

\paragraph{Hardware:} This experiment was performed on an Intel Xeon Gold 6148 CPU.

\clearpage

\section{SFA for reversible processes} 
\label{sec:rev}

The update for $\W$ in Alg.~\ref{alg:online} requires the input neurons to store both the input, $\x_t$, and the delayed sum, $\bar\x_t$.
Here, we propose a modification of the algorithm, which is exactly SFA in the case that the expanded input $\{\x_t\}$ is reversible, that only requires the input neurons to store the input $\x_t$.
Suppose the expanded signal $\{\x_t\}$ exhibits time-reversal symmetry; that is,
\begin{align*}
    \frac1T\sum_{t=1}^T\x_t\x_{t-1}^\top=\frac1T\sum_{t=1}^T\x_{t-1}\x_t^\top.
\end{align*}
Then
\begin{align*}
    \C_{\bar x\bar x}&=\frac1T\sum_{t=1}^T\bar\x_t(\x_t+\x_{t-1})^\top\\
    &=\frac1T\sum_{t=1}^T\bar\x_t\x_t^\top+\frac1T\sum_{t=1}^T(\x_t+\x_{t-1})\x_{t-1}^\top\\
    &=\frac1T\sum_{t=1}^T\bar\x_t\x_t^\top+\frac1T\sum_{t=1}^T\x_{t-1}\x_t^\top+\frac1T\sum_{t=1}^T\x_t\x_t^\top+\frac1T(\x_0\x_0^\top-\x_T\x_T^\top)\\
    &=\frac2T\sum_{t=1}^T\bar\x_t\x_t^\top+\frac1T(\x_0\x_0^\top-\x_T\x_T^\top)\\
    &=2\C_{\bar xx}+\frac1T(\x_0\x_0^\top-\x_T\x_T^\top),
\end{align*}
where 
\begin{align*}
    \C_{\bar xx}:=\frac1T\sum_{t=1}^T\bar\x_t\x_t^\top.
\end{align*}
In the large $T$ limit, we can approximate the offline gradient descent update for $\W$ in Bio-SFA by replacing $\C_{\bar x\bar x}$ with $2\C_{\bar xx}$, which results in the update
\begin{align*}
    \W\gets\W+2\eta(2\M^{-1}\W\C_{\bar xx}-\W\C_{xx}).
\end{align*}
Recalling that $\bar\y_t=\M^{-1}\W\bar\x_t$, we can write the online stochastic gradient descent step for $\W$ as
\begin{align*}
    \W&\gets\W+2\eta(2\bar\y_t-\a_t)\x_t^\top.
\end{align*}
This yields our online SFA algorithm for reversible processes (Alg.~\ref{alg:online_rev}).

\begin{algorithm}[ht]
  \caption{Bio-SFA for reversible processes}
  \label{alg:online_rev}
\begin{algorithmic}
  \STATE {\bfseries input} expanded signal $\{\x_0,\x_1,\dots,\x_T\}$; dimension $k$; parameters $\gamma$, $\eta$, $\tau$
  \STATE {\bfseries initialize} matrix $\W$ and positive definite matrix $\M$
  \FOR{$t=1,2,\dots,T$}
  \STATE $\a_t\gets\W\x_t$
  \REPEAT
  \STATE $\y_t\gets\y_t+\gamma(\a_t-\M\y_t)$ \hfill $\triangleright\;$ compute output
  \UNTIL{convergence}
  \STATE $\bar\x_t\gets \x_t+\x_{t-1}$
  \STATE $\bar\y_t\gets\y_t+\y_{t-1}$ 
  \STATE $\W \gets \W + 2\eta (\bar\y_t - \a_t)\x_t^\top$ \hfill $\triangleright\;$ stochastic gradient descent-ascent steps
  \STATE $\M \gets \M + \frac{\eta}{\tau} (\bar\y_t\bar\y_t^\top-\M) $
  \ENDFOR
\end{algorithmic}
\end{algorithm}

\begin{figure}[ht!]
\centering
\makebox[0pt][c]{\parbox{\textwidth}{%
    \begin{minipage}[c]{0.5\textwidth}\centering
    \resizebox{1\textwidth}{!}{
        \includegraphics{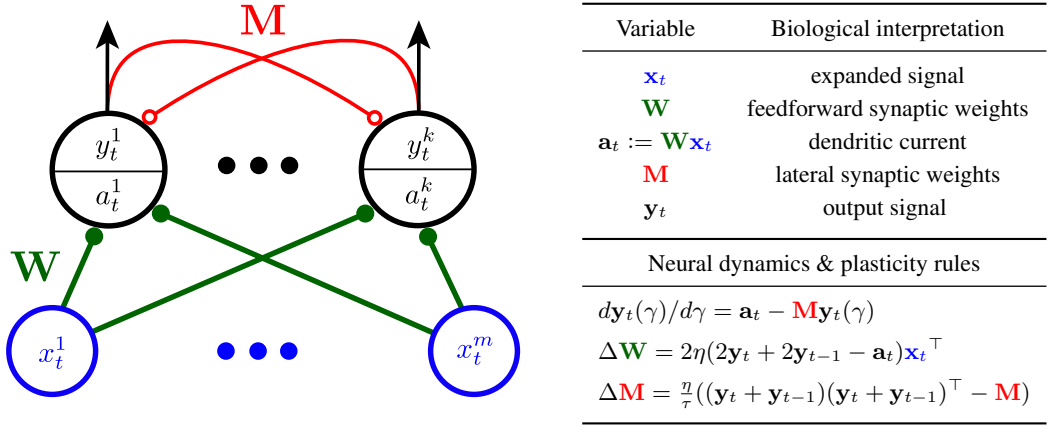}
        }
    \end{minipage}
    \hfill
    \begin{minipage}[c]{0.46\textwidth}\centering
    \resizebox{1\textwidth}{!}{
        \renewcommand{\arraystretch}{1.25}
        \small
        \begin{tabular}{  c  c }
            \toprule
            Variable & Biological interpretation \\
            \midrule
            ${\color{blue}\x_t}$ & expanded signal \\ 
            ${\color{green}\W}$ & feedforward synaptic weights \\ 
            $\a_t:={\color{green}\W}{\color{blue}\x_t}$ & dendritic current \\ 
            ${\color{red}\M}$ & lateral synaptic weights \\ 
            $\y_t$ & output signal
            \vspace{5pt}
            \\ 
            \toprule
            \multicolumn{2}{c}{Neural dynamics \& plasticity rules}\\
            \midrule
            \multicolumn{2}{l}{$d\y_t(\gamma)/d\gamma=\a_t-{\color{red}\M}\y_t(\gamma)$}\vspace{3pt}\\ 
            \multicolumn{2}{l}{$\Delta{\color{green}\W}=2\eta(2\y_t+2\y_{t-1}-\a_t){\color{blue}\x_t}^\top$}\vspace{3pt}\\
            \multicolumn{2}{l}{$\Delta{\color{red}\M}=\frac{\eta}{\tau}((\y_t+\y_{t-1})(\y_t+\y_{t-1})^\top-{\color{red}\M})$}
            \vspace{3pt}\\
        \bottomrule
        \end{tabular}
        }
    \end{minipage}%
}}
\caption{A biologically plausible neural network implementation of Bio-SFA for reversible processes. The figure on the left depicts the architecture of the neural network. Blue circles are the input neurons and black circles are the output neurons with separate dendritic and somatic compartments. Lines with circles connecting the neurons denote synapses. Filled (resp.\ empty) circles denote excitatory (resp.\ inhibitory) synapses.}
\label{fig:NN_supp}
\end{figure}

As with Bio-SFA, Alg.~\ref{alg:online_rev} can be implemented in the neural network shown in Fig.~\ref{fig:NN_supp}.
Note that in this case, the elementwise synaptic update for $W_{ij}$, given by
\begin{align*}
    W_{ij}\gets W_{ij}+2\eta(2\bar y_t^i-a_t^i)x_t^j,
\end{align*}
depends only on $\bar y_t^i$, $a_t^i$ and $x_t^j$, so the pre-synaptic input neuron only needs to represent the $x_t^j$, as opposed to both $x_t^j$ and $\bar x_t^j$.
Biologically, this is more realistic because the signal frequency of dendrites is slower than the signal frequency of axons, so it is more likely that slow variables are represented in the post-synaptic neuron.

\begin{figure}[ht]
\centering
\subfloat[Optimal Stimuli]{\label{fig:filters_rev}\includegraphics[height=0.48\textwidth]{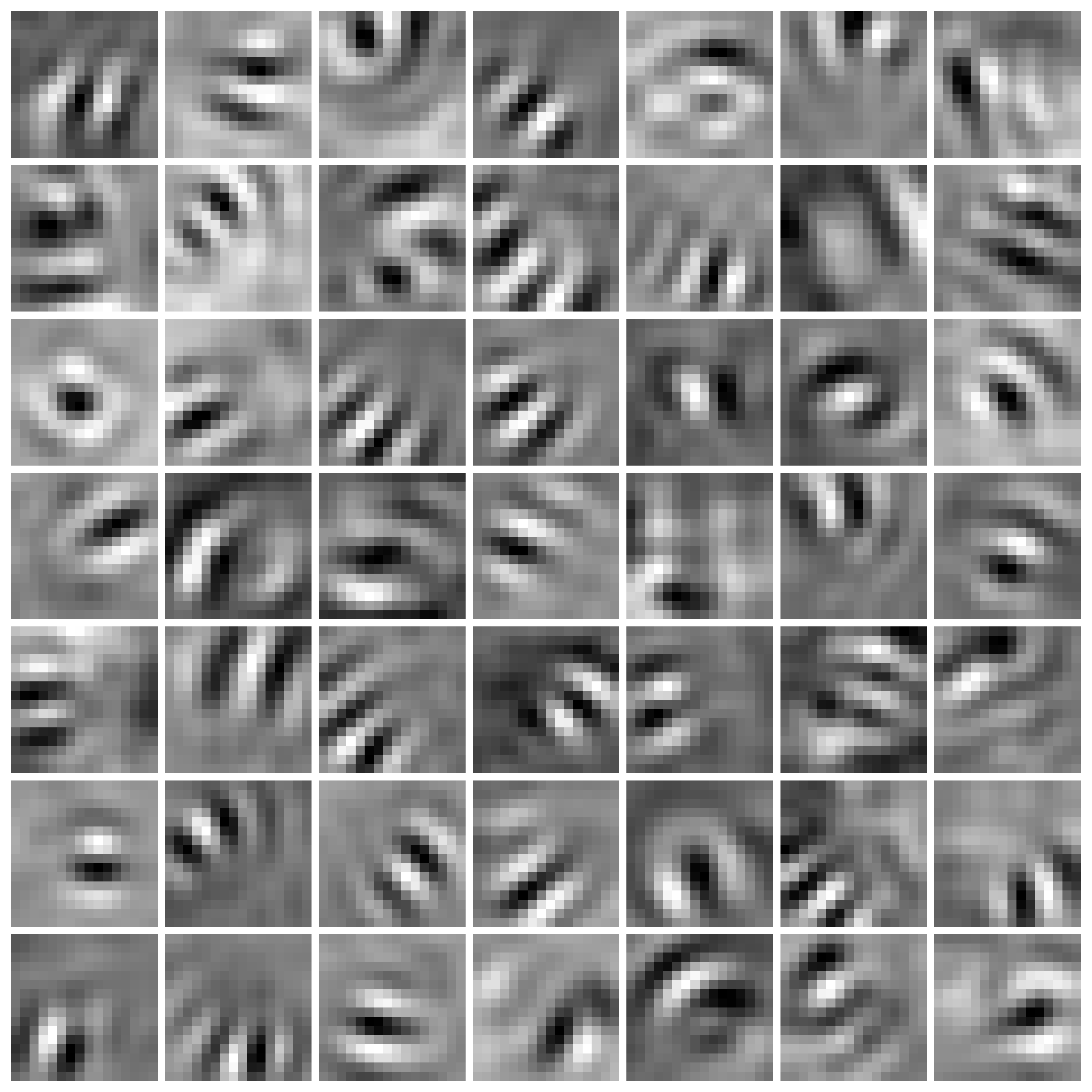}}
\hfill
\subfloat[Error]{\label{fig:convergence_supp}\includegraphics[height=0.48\textwidth]{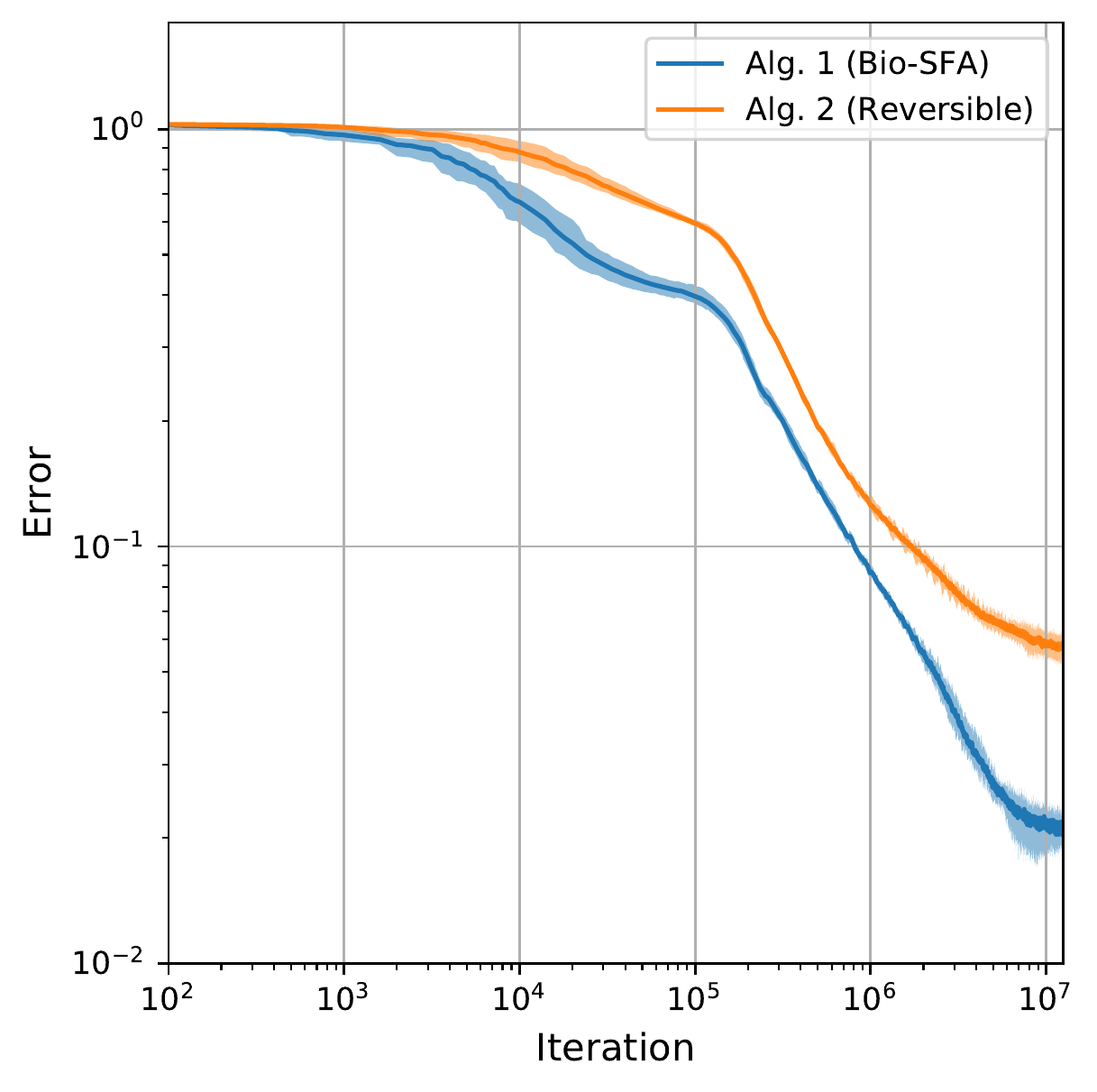}}
\caption{Performance of Bio-SFA for reversible processes on a sequence of natural images. Panel (a) shows the maximally excitatory stimuli for the 49-dimensional output obtained by Bio-SFA for reversible processes. Panel (b) shows the mean error and 90\% confidence intervals over 10 runs.}
\end{figure}

We test Alg.~\ref{alg:online_rev} on the naturalistic image sequences from \cite{vanHateren1998}, which are not reversible (due to the rotation of the images). 
In Fig.~\ref{fig:filters_rev}, we display the optimal stimuli for the filters that are found by Alg.~\ref{alg:online_rev}.
These optimal stimuli are in close qualitative agreement with the optimal stimuli found by Bio-SFA, shown in Fig.~\ref{fig:filters}.
In Fig.~\ref{fig:convergence_supp}, we plot the error defined in Eq.~\eqref{eq:error} and find that Alg.~\ref{alg:online_rev} (Bio-SFA for reversible processes) performs comparably with Alg.~\ref{alg:online} (Bio-SFA).

\end{document}

%% file: commands.tex
\theoremstyle{definition}

\DeclareMathOperator{\tr}{Tr}

\newcommand{\argmax}[1]{\underset{#1}{\operatorname{arg}\operatorname{max}}\;}
\newcommand{\argmin}[1]{\underset{#1}{\operatorname{arg}\operatorname{min}}\;}

\renewcommand{\a}{{\bf a}}

\newcommand{\s}{{\bf s}}

\newcommand{\x}{{\bf x}}
\newcommand{\y}{{\bf y}}
\newcommand{\z}{{\bf z}}

\newcommand{\C}{{\bf C}}

\newcommand{\I}{{\bf I}}

\newcommand{\M}{{\bf M}}

\newcommand{\R}{\mathbb{R}}

\newcommand{\U}{{\bf U}}
\newcommand{\V}{{\bf V}}
\newcommand{\W}{{\bf W}}

\newcommand{\X}{{\bf X}}
\newcommand{\Y}{{\bf Y}}

\renewcommand{\v}{{\bf v}}

\newcommand{\Lam}{\boldsymbol{\Lambda}}

\mathchardef\ordinarycolon\mathcode`\:
\mathcode`\:=\string"8000
\begingroup \catcode`\:=\active
  \gdef:{\mathrel{\mathop\ordinarycolon}}
\endgroup

%% file: main.bbl
\begin{thebibliography}{10}

\bibitem{berkes2002applying}
Pietro Berkes and Laurenz Wiskott.
\newblock Applying slow feature analysis to image sequences yields a rich
  repertoire of complex cell properties.
\newblock In {\em International Conference on Artificial Neural Networks},
  pages 81--86. Springer, 2002.

\bibitem{berkes2005slow}
Pietro Berkes and Laurenz Wiskott.
\newblock Slow feature analysis yields a rich repertoire of complex cell
  properties.
\newblock {\em Journal of Vision}, 5(6):9--9, 2005.

\bibitem{BlaschkeBerkesEtAl-2006}
T.~Blaschke, P.~Berkes, and L.~Wiskott.
\newblock What is the relationship between slow feature analysis and
  independent component analysis?
\newblock {\em Neural Computation}, 18(10):2495--2508, 2006.

\bibitem{blaschke2004cubica}
T.~{Blaschke} and L.~{Wiskott}.
\newblock Cubica: independent component analysis by simultaneous third- and
  fourth-order cumulant diagonalization.
\newblock {\em IEEE Transactions on Signal Processing}, 52(5):1250--1256, 2004.

\bibitem{clark2019unsupervised}
David Clark, Jesse Livezey, and Kristofer Bouchard.
\newblock Unsupervised discovery of temporal structure in noisy data with
  dynamical components analysis.
\newblock In {\em Advances in Neural Information Processing Systems}, pages
  14267--14278, 2019.

\bibitem{cox2000multidimensional}
Trevor~F Cox and Michael~AA Cox.
\newblock {\em Multidimensional Scaling}.
\newblock Chapman and Hall/CRC, 2000.

\bibitem{creutzig2008predictive}
Felix Creutzig and Henning Sprekeler.
\newblock Predictive coding and the slowness principle: An
  information-theoretic approach.
\newblock {\em Neural Computation}, 20(4):1026--1041, 2008.

\bibitem{Fldik1991}
Peter F\"{o}ldi{\'{a}}k.
\newblock Learning invariance from transformation sequences.
\newblock {\em Neural Computation}, 3(2):194--200, June 1991.

\bibitem{franzius2007hsfa}
Mathias Franzius, Henning Sprekeler, and Laurenz Wiskott.
\newblock Slowness and sparseness lead to place, head-direction, and
  spatial-view cells.
\newblock {\em {PLoS} Computational Biology}, 3(8):e166, 2007.

\bibitem{Hyvrinen2000}
A.~Hyv\"{a}rinen and E.~Oja.
\newblock Independent component analysis: algorithms and applications.
\newblock {\em Neural Networks}, 13(4-5):411--430, June 2000.

\bibitem{hyvarinen1999fast}
Aapo Hyvarinen.
\newblock Fast and robust fixed-point algorithms for independent component
  analysis.
\newblock {\em IEEE transactions on Neural Networks}, 10(3):626--634, 1999.

\bibitem{hyvarinen2000independent}
Aapo Hyv{\"a}rinen and Erkki Oja.
\newblock Independent component analysis: Algorithms and applications.
\newblock {\em Neural Networks}, 13(4-5):411--430, 2000.

\bibitem{koch1992multiplying}
Christof Koch and Tomaso Poggio.
\newblock Multiplying with synapses and neurons.
\newblock In {\em Single neuron computation}, pages 315--345. Elsevier, 1992.

\bibitem{kompella2012incremental}
Varun~Raj Kompella, Matthew Luciw, and J{\"u}rgen Schmidhuber.
\newblock Incremental slow feature analysis: Adaptive low-complexity slow
  feature updating from high-dimensional input streams.
\newblock {\em Neural Computation}, 24(11):2994--3024, 2012.

\bibitem{lipshutz2020biologically}
David Lipshutz, Yanis Bahroun, Siavash Golkar, Anirvan~M. Sengupta, and
  Dmitri~B. Chkovskii.
\newblock A biologically plausible neural network for multi-channel canonical
  correlation analysis.
\newblock {\em arXiv preprint arXiv:2010.00525}, 2020.

\bibitem{liwicki2012iksfa}
Stephan Liwicki, Stefanos Zafeiriou, and Maja Pantic.
\newblock Incremental slow feature analysis with indefinite kernel for online
  temporal video segmentation.
\newblock In {\em Computer Vision -- ACCV 2012}, volume 7725, pages 162--176.
  Springer Berlin Heidelberg, 2013.

\bibitem{Malik2014}
Zeeshan~Khawar Malik, Amir Hussain, and Jonathan Wu.
\newblock Novel biologically inspired approaches to extracting online
  information from temporal data.
\newblock {\em Cognitive Computation}, 6(3):595--607, April 2014.

\bibitem{mel1990sigma}
Bartlett~W Mel and Christof Koch.
\newblock Sigma-pi learning: On radial basis functions and cortical associative
  learning.
\newblock In {\em Advances in Neural Information Processing Systems}, pages
  474--481, 1990.

\bibitem{mitchison1991removing}
Graeme Mitchison.
\newblock Removing time variation with the anti-{H}ebbian differential synapse.
\newblock {\em Neural Computation}, 3(3):312--320, 1991.

\bibitem{No2015}
Frank No{\'{e}} and Cecilia Clementi.
\newblock Kinetic distance and kinetic maps from molecular dynamics simulation.
\newblock {\em Journal of Chemical Theory and Computation}, 11(10):5002--5011,
  September 2015.

\bibitem{pehlevan2015normative}
Cengiz Pehlevan and Dmitri Chklovskii.
\newblock A normative theory of adaptive dimensionality reduction in neural
  networks.
\newblock In {\em Advances in Neural Information Processing Systems}, pages
  2269--2277, 2015.

\bibitem{PrezHernndez2013}
Guillermo P{\'{e}}rez-Hern{\'{a}}ndez, Fabian Paul, Toni Giorgino, Gianni~De
  Fabritiis, and Frank No{\'{e}}.
\newblock Identification of slow molecular order parameters for {M}arkov model
  construction.
\newblock {\em The Journal of Chemical Physics}, 139(1):015102, July 2013.

\bibitem{rumelhart1986general}
David~E Rumelhart, Geoffrey~E Hinton, James~L McClelland, et~al.
\newblock A general framework for parallel distributed processing.
\newblock {\em Parallel distributed processing: Explorations in the
  microstructure of cognition}, 1(45-76):26, 1986.

\bibitem{schonfeld2013ratlab}
Fabian Sch\"{o}nfeld and Laurenz Wiskott.
\newblock {RatLab}: an easy to use tool for place code simulations.
\newblock {\em Frontiers in Computational Neuroscience}, 7, 2013.

\bibitem{schonfeld2015hsfa}
Fabian Sch{\"o}nfeld and Laurenz Wiskott.
\newblock Modeling place field activity with hierarchical slow feature
  analysis.
\newblock {\em Frontiers in Computational Neuroscience}, 9, 2015.

\bibitem{Schwantes2015}
Christian~R. Schwantes and Vijay~S. Pande.
\newblock Modeling molecular kinetics with {tICA} and the kernel trick.
\newblock {\em Journal of Chemical Theory and Computation}, 11(2):600--608,
  January 2015.

\bibitem{MSultan2017}
Mohammad~M. Sultan and Vijay~S. Pande.
\newblock {tICA}-metadynamics: Accelerating metadynamics by using kinetically
  selected collective variables.
\newblock {\em Journal of Chemical Theory and Computation}, 13(6):2440--2447,
  May 2017.

\bibitem{vanHateren1998}
J.~H. van Hateren and A.~van~der Schaaf.
\newblock Independent component filters of natural images compared with simple
  cells in primary visual cortex.
\newblock {\em Proceedings of the Royal Society of London. Series B: Biological
  Sciences}, 265(1394):359--366, 1998.

\bibitem{weghenkel2018slowness}
Bj{\"o}rn Weghenkel and Laurenz Wiskott.
\newblock Slowness as a proxy for temporal predictability: An empirical
  comparison.
\newblock {\em Neural computation}, 30(5):1151--1179, 2018.

\bibitem{wiskott2003estimating}
Laurenz Wiskott.
\newblock Estimating driving forces of nonstationary time series with slow
  feature analysis.
\newblock {\em arXiv preprint cond-mat/0312317}, 2003.

\bibitem{wiskott2002slow}
Laurenz Wiskott and Terrence~J Sejnowski.
\newblock Slow feature analysis: {U}nsupervised learning of invariances.
\newblock {\em Neural Computation}, 14(4):715--770, 2002.

\bibitem{Yousefi2012}
Bardia Yousefi and Chu~Kiong Loo.
\newblock Development of fast incremental slow feature analysis (f-{IncSFA}).
\newblock In {\em The 2012 International Joint Conference on Neural Networks
  ({IJCNN})}. {IEEE}, June 2012.

\bibitem{QingfuZhang2000}
Qingfu Zhang and Yiu~Wing Leung.
\newblock A class of learning algorithms for principal component analysis and
  minor component analysis.
\newblock {\em {IEEE} Transactions on Neural Networks}, 11(1):200--204, 2000.

\bibitem{zhang2012slow}
Zhang Zhang and Dacheng Tao.
\newblock Slow feature analysis for human action recognition.
\newblock {\em IEEE Transactions on Pattern Analysis and Machine Intelligence},
  34(3):436--450, 2012.

\bibitem{Zito2008mdp}
Tiziano Zito.
\newblock Modular toolkit for data processing ({MDP}): a python data processing
  framework.
\newblock {\em Frontiers in Neuroinformatics}, 2, 2008.

\end{thebibliography}
